 \documentclass[aps,twocolumn,showpacs,showkeys,10pt]{revtex4}

  \usepackage{graphicx}
  \usepackage{dcolumn}
  \usepackage{bm}

 \newcommand{\ud}{\mathrm{d}}
 \newcommand{\eqn}[1]{(\ref{#1})}
 \newcommand{\fig}[1]{Fig.~\ref{#1}}
 \newcommand{\figs}[1]{Figs.~\ref{#1}}
 \newcommand{\Fig}[1]{Figure~\ref{#1}}

\begin{document}

 \title{Statistical Properties of Nonlinear Phase Noise}

 \author{Keang-Po Ho}

 \affiliation{Institute of Communications Engineering and Department of Electrical Engineering\\ National Taiwan University\\ No. 1, Sec. 4, Roosevelt Rd., Taipei 106, Taiwan}
 \email{kpho@cc.ee.ntu.edu.tw}

 \date{\today}%
 \begin{abstract}
 The statistical properties of nonlinear phase noise, often called the Gordon-Mollenauer effect, is studied analytically when the number of fiber spans is very large.
 The joint characteristic functions of the nonlinear phase noise with electric field, received intensity, and the phase of amplifier noise are all derived analytically.
 Based on the joint characteristic function of nonlinear phase noise with the phase of amplifier noise, the error probability of signal having nonlinear phase noise is calculated using the Fourier series expansion of the probability density function.
 The error probability is increased due to the dependence between nonlinear phase noise and the phase of amplifier noise.
 When the received intensity is used to compensate the nonlinear phase noise, the optimal linear and nonlinear minimum mean-square error compensators are derived analytically using the joint characteristic function of nonlinear phase noise and received intensity.
Using the joint probability density of received amplitude and phase, the optimal maximum {\em a posteriori} probability detector is derived analytical.
 The nonlinear compensator always performs better than linear compensator.
 \end{abstract}



 \maketitle


 \vspace{1cm}
 {\bf Revision History}

 \begin{tabular}{|l|l|}
 \hline
 Date & Revisions \\
 \hline
 Mar 03 & Initial draft, to be published. \\
 May 03 & Additional references \\
   & Add to Sec. \ref{sec:ber}, submitted to JLT \\
Jun 03 & Add Sec. \ref{sec:cfPhiYTheta} \\
        & Change Sec. \ref{sec:lincomp}, submitted to JLT \\
Dec 03 & Add Sec. \ref{sec:nlcomp}, submitted to JLT \\
\hline
 \end{tabular}


 \section{Introduction}

 When optical amplifiers are used to compensate for fiber loss, the interaction of amplifier noises and the fiber Kerr effect causes nonlinear phase noise, often called the Gordon-Mollenauer effect \cite{gordon90}, or more precisely, self-phase modulation induced nonlinear phase noise.
 Nonlinear phase noise degrades phase-modulated signal like phase-shifted keying (PSK) and differential phase-shift keying (DPSK) signal \cite{gordon90, ryu92, saito93, mecozzi94, mckinstrie02, kim03, xu03, ho0309a, ho0309b, kim03xpm, mizuochi03, wei03spm, ho0403b}.
 This class of constant-intensity modulation has renewed attention recently for long haul and/or spectral efficiency transmission systems \cite{gnauck02, griffin02, zhu02, miyamoto02, bissessur03, gnauck03, cho03, rasmussen03, zhu03, vareille03, tsuritani03, cai03, gnauck03tb}, mostly DPSK signal using return-to-zero (RZ) pulses or differential quadrature phase-shift keying (DQPSK) signal \cite{griffin02, griffin02a, cho03, griffin03, kim03a, wree03a}.
 The comparison of DPSK to on-off keying signal shows advantage of DPSK signal in certain applications \cite{hoshida02, leibrich02, xu03, mizuochi03}.

 Traditionally, the performance of a system with nonlinear phase noise is evaluated based on the phase variance \cite{gordon90, mckinstrie02, liu02, xu02, ho0403a, mckinstrie02a, xu03} or spectral broadening \cite{ryu92, saito93, mecozzi94, mizuochi03}. 
 However, it is found that the nonlinear phase noise is not Gaussian-distributed both experimentally \cite{kim03} and analytically \cite{mecozzi94, ho0309c, ho0308}.  
 For non-Gaussian noise, neither the variance nor the $Q$-factor \cite{wei03, wei03spm} is sufficient to characterize the performance of the system.
 The probability density function (p.d.f.) is necessary to better understand the noise properties and evaluates the system performance. 

 The p.d.f.~of nonlinear phase noise alone \cite{ho0309c, ho0308} is not sufficient to characterize the signal with nonlinear phase noise.
 Because of the dependence between nonlinear phase noise and signal phase, the joint p.d.f.~of the nonlinear phase noise and the signal phase is necessary to find the error probability for a phase-modulated signal.  
 This article provides analytical expressions of the joint asymptotic characteristic functions of the nonlinear phase noise and the received electric field without nonlinear phase noise.  
 The amplifier noise is asymptotically modeled as a distributed process for a large number of fiber spans.  
 After the characteristic function is derived analytically, the p.d.f.~is the inverse Fourier transform of the corresponding characteristic function.
 The dependence between nonlinear phase noise and the phase of amplifier noise increases the error probability.

 The received phase is the summation of the nonlinear phase noise and the phase of amplifier noise.
Although it is obvious that nonlinear phase noise is uncorrelated with the phase of amplifier noise \cite{ho0309a, ho0309b}, as non-Gaussian random variables, they are weakly depending on each other.
 Using the joint characteristic function of nonlinear phase noise and the phase of amplifier noise, the p.d.f.~of the received phase can be expanded as a Fourier series.
 Using the Fourier series, the error probability of PSK and DPSK signal is evaluated by a series summation.
 Because the nonlinear phase noise has a weak dependence on the phase of amplifier noise, the Fourier series expansion is more complicated than traditional method in which the extra phase noise is independent of the signal phase \cite{jain74, nicholson84} or the approximation of \cite{ho0309a, ho0309b}.
For PSK signals, in contrary to \cite{mecozzi94}, the received phase does not distribute symmetrically with respect to the mean nonlinear phase shift.

 Correlated with each other, the received intensity can be used to compensate the nonlinear phase noise.
 When a linear compensator compensates the nonlinear phase noise using a scaled version of the received intensity \cite{ho0403a, liu02, xu02, xu02a}, the optimal linear compensator to minimize the variance of the residual nonlinear phase noise is found using the joint characteristic function of nonlinear phase noise and received intensity.
However, as the nonlinear phase noise is not Gaussian distributed, the minimum mean-square linear compensator does not necessary minimize the error probability of the compensated signals.
When the exact error probability of linearly compensated signals is derived, the optimal linear compensator can be found using numerical optimization.

 The minimum mean-square error (MMSE) compensator is the conditional mean of the nonlinear phase noise given the received intensity  \cite{ho0306} \cite[Sec. 10.2]{mcdonough2}.
 Using the conditional characteristic function of the nonlinear phase noise given the received intensity, the optimal nonlinear compensator is found to perform slightly better than the linear compensator.
The joint p.d.f.~of the received amplitude and phase can also express as a Fourier series with Fourier coefficients depending on the received amplitude.
Using the joint p.d.f.~of the received amplitude and phase, the optimal detector and the corresponding compensator can be derived to minimize the error probability of a PSK signal with nonlinear phase noise.

Although very popular, MMSE compensator does not minimize the error probability after the compensator. 
To minimize the error probability, the optimal maximum {\em a posteriori} probability (MAP) detector \cite[Sec. 5.8]{mcdonough2} must be used.
It is theoretically very important to find the optimal compensator, possible by any mean and regardless of complexity or practicality, to combat nonlinear phase noise.
An application of the optimal MAP compensator is to verify the optimality of a practical compensator.
In order to find the optimal nonlinear MAP compensator, this paper first derives the joint distribution of the received amplitude and phase. 
The optimal nonlinear MAP compensator is then derived for PSK signal with nonlinear phase noise.
The error probability of both nonlinear MAP and MMSE compensators is calculated using the joint distribution of received amplitude and phase.

 Later parts of this paper are organized as following:
 Sec. \ref{sec:basic} builds the mathematical model of the nonlinear phase noise and derives the joint characteristic function of the normalized nonlinear phase noise and the electric field with nonlinear phase noise.
 Sec. \ref{sec:non} gives the marginal p.d.f of nonlinear phase noise by inverse Fourier transform.
 Sec. \ref{sec:jcf} obtains the joint characteristic functions of the nonlinear phase noise with received intensity and/or the phase of amplifier noise.
 Using the joint characteristic function of nonlinear phase noise with the phase of amplifier noise, Sec. \ref{sec:ber} calculates the exact error probability of PSK and DPSK signals with nonlinear phase noise.
 An approximation is also presented based on the assumption that the nonlinear phase noise is independent of the phase of amplifier noise. 
 Using the joint characteristic function of nonlinear phase noise with received intensity, Sec. \ref{sec:lincomp} provides the optimal linear compensators to compensate the nonlinear phase noise using received intensity.
Using the joint characteristic function of nonlinear phase noise with both received intensity and phase of amplifier noise, the exact error probability of PSK and DPSK signals with linearly compensated nonlinear phase is derived.
Sec. \ref{sec:nlcomp} discusses the nonlinear compensator for nonlinear phase noise to minimize either the error probability or the variance of residual nonlinear phase noise.
 Finally, Sec. \ref{sec:end} is the conclusion of this article. 

 \section{Joint Statistics of Nonlinear Phase Noise and Electric Field}
 \label{sec:basic}

 This section provides the joint characteristic function of nonlinear phase noise and the electric field without nonlinear phase noise.
 Both the nonlinear phase noise and the electric field are first normalized and represented as the summation of infinite number of independently distributed random variables.
 The joint characteristic function of nonlinear phase noise and electric field is the product of the corresponding joint characteristic functions of those random variables.
 After some algebraic simplifications, the joint characteristic function has a simple expression.

 \subsection{Normalization of Nonlinear Phase Noise}

 In a lightwave system, nonlinear phase noise is induced by the interaction of fiber Kerr effect and optical amplifier noise \cite{gordon90}.
 In this article, nonlinear phase noise is induced by self-phase modulation through the amplifier noise in the same polarization as the signal and within an optical bandwidth matched to the signal. 
 The phase noise induced by cross-phase modulation from amplifier noise outside that optical bandwidth is ignored for simplicity.
 The amplifier noise from the orthogonal polarization is also ignored for simplicity.
 As shown later, we can include the phase noise from cross-phase modulation or orthogonal polarization by simple modification.

 For an $N$-span fiber system, the overall nonlinear phase noise is \cite{gordon90, ho0403a, ho0309c, liu02, ho0306}

 \begin{eqnarray}
 \phi_{\mathrm{NL}}&=&\gamma L_{\mathrm{eff}} \left\{|\vec{E}_0 + \vec{n}_1|^2 +
	  |\vec{E}_0 +\vec{n}_1 + \vec{n}_2|^2 + \cdots \right. \nonumber \\
 &&\left. + |\vec{E}_0 + \vec{n}_1 + \cdots + \vec{n}_N|^2 \right\},
 \label{phiNL}
 \end{eqnarray}

 \noindent where $\vec{E}_0$ is a two-dimensional vector as the baseband representation of the transmitted electric field, $\vec{n}_k, k = 1, \dots, N$, are independent identically distributed (i.i.d.) zero-mean circular Gaussian random vectors as the optical amplifier noise introduced into the system at the $k^{\mathrm{th}}$ fiber span, $\gamma L_{\mathrm{eff}}$ is the product of fiber nonlinear coefficient and the effective fiber length per span. 
 In \eqn{phiNL}, both electric field of $\vec{E}_0$ and amplifier noises of $\vec{n}_k$ can also be represented as a complex number.

 \begin{figure*}
 \centerline{
  \begin{tabular}{ccc}
	 {\includegraphics[width = 0.4 \textwidth]{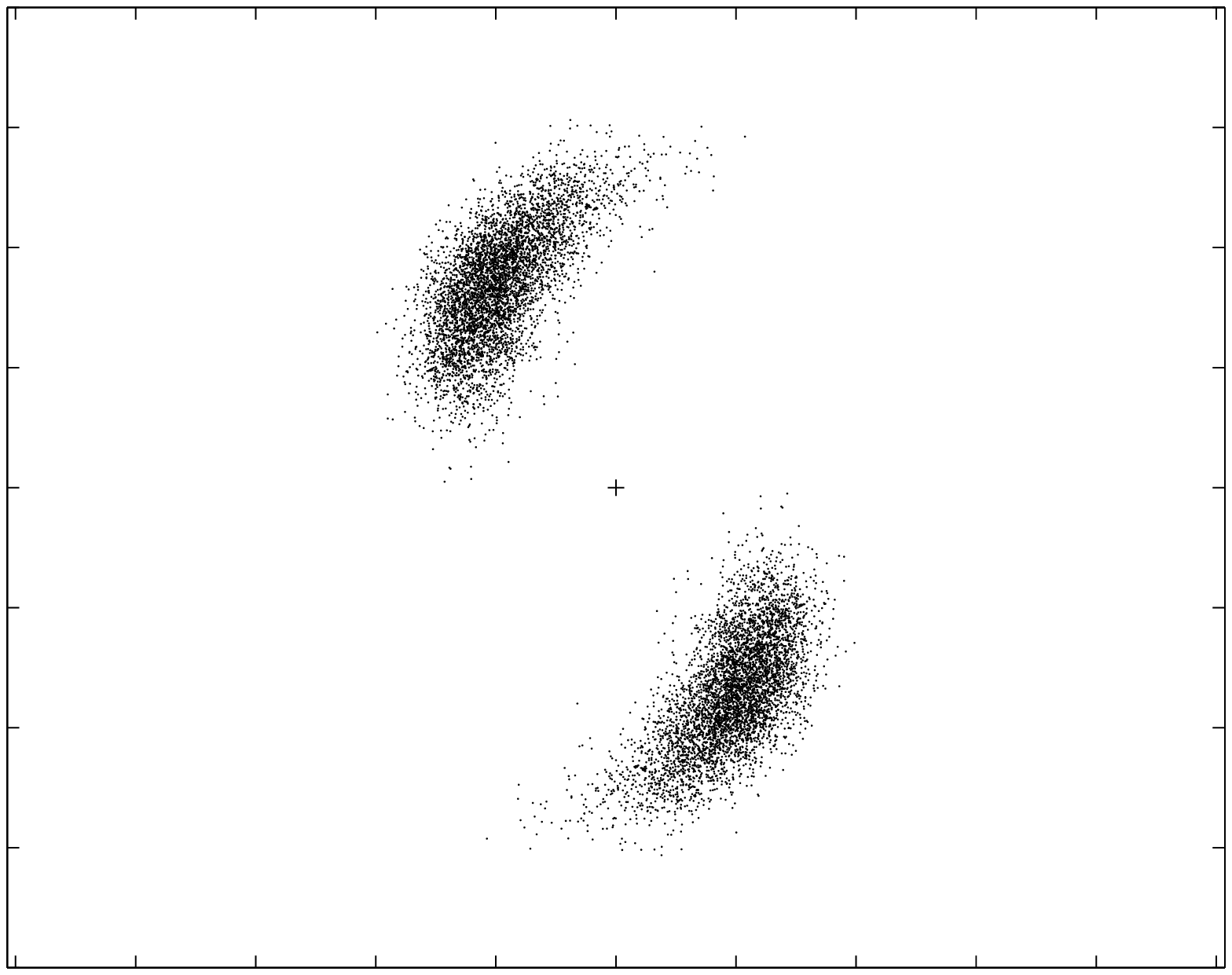}} & \hspace{.25cm} & {\includegraphics[width = 0.4 \textwidth]{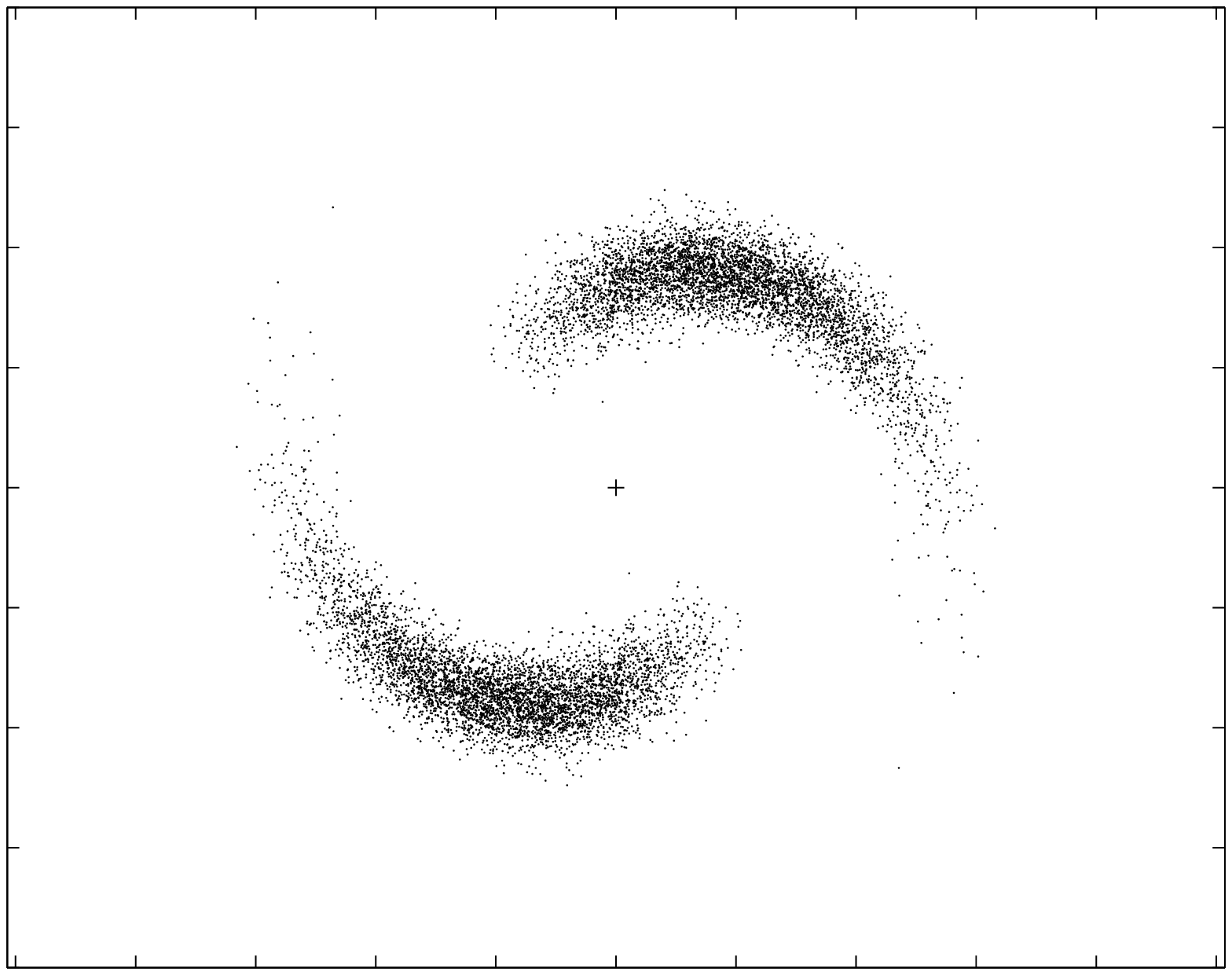}} \\
	 (a) $<\!\!\Phi_{\mathrm{NL}}\!\!> = 1$ rad & \hspace{.25cm} & (b) $<\!\!\Phi_{\mathrm{NL}}\!\!> = 2$ rad 
  \end{tabular}
 }
 \caption{Simulated distribution of the received electric field for mean nonlinear phase shift of (a) $<\!\!\Phi_{\mathrm{NL}}\!\!> = 1$ rad and (b) $<\!\!\Phi_{\mathrm{NL}}\!\!> = 2$ rad. }
 \label{figyinyang}
 \end{figure*}

 \figs{figyinyang} show the simulated distribution of the received electric field including the contribution from nonlinear phase noise. 
 The mean nonlinear phase shifts $<\!\!\Phi_{\mathrm{NL}}\!\!>$  are 1 and 2 rad for Figs. \ref{figyinyang}a and \ref{figyinyang}b, respectively. 
 The mean nonlinear phase shift of $<\!\!\Phi_{\mathrm{NL}}\!\!> = 1$ rad corresponds to the limitation estimated by \cite{gordon90}.
 The mean nonlinear phase shift of $<\!\!\Phi_{\mathrm{NL}}\!\!> = 2$ rad corresponds to the limitation given by \cite{ho0403a} when the standard deviation of nonlinear phase noise is halved using a linear compensator. 
 The limitation of $<\!\!\Phi_{\mathrm{NL}}\!\!> = 2$ rad may be inferred from \cite{xu02, liu02}.

 \figs{figyinyang} are plotted for the case that the signal-to-noise ratio (SNR) $\rho_s = 18$ (12.6 dB), corresponding to an error probability of $10^{-9}$ if the amplifier noise is the sole impairment. 
 The number of spans is $N = 32$. 
 The transmitted signal is $\vec{E_0} = (\pm|\vec{E_0}|, 0)$ for binary PSK signal. 
 The distribution of  Figs. \ref{figyinyang} has $5000$ points for different noise combinations. 
 In practice,  the signal distribution of \figs{figyinyang} can be measured using an optical phase-locked loop (see Fig. 5 of \cite{norimatsu92}). 
 Note that although the optical phase-locked loop actually tracks out the mean nonlinear phase shift of $<\!\!\Phi_{\mathrm{NL}}\!\!>$, nonzero values of $<\!\!\Phi_{\mathrm{NL}}\!\!>$ have been preserved in plotting \figs{figyinyang} to better illustrate the nonlinear phase noise. 

 With large number of fiber spans, the summation of \eqn{phiNL} can be replaced by integration as \cite{ho0308, liu02}

 \begin{equation}
 \phi_{\mathrm{NL}} = \kappa \int_{0}^{L} |\vec{E}_0 + \vec{S}(z)|^2 \ud z,
 \label{phiint}
 \end{equation} 

 \noindent where $L$ is the overall fiber length, $\kappa = N \gamma  L_{\mathrm{eff}}/L$ is the average nonlinear coefficient per unit length, and $\vec{S}(z)$ is a zero-mean two-dimensional  Brownian motion of $E\{\vec{S}(z_1) \cdot \vec{S}(z_2)\} = \sigma^2_{s} \min(z_1, z_2)$, where $\cdot$ denotes inner product of two vectors.
 The variance of $\sigma_s^2 = N \sigma^2_{\mathrm{ASE}}/L$ is the noise variance per unit length where $E\{|\vec{n}_k|^2\} = \sigma^2_{\mathrm{ASE}}, k = 1, \ldots, N $ is noise variance per amplifier per polarization in the optical bandwidth matched to the signal.

 In this article, we investigate the joint statistical properties of the normalized electric field and normalized nonlinear phase noise

 \begin{equation}
 \vec{e}_N = \vec{\xi}_0 + \vec{b}(1), \quad \phi = \int_{0}^{1} |\vec{\xi}_0 + \vec{b}(t)|^2 \ud t,
 \label{phiNor}
 \end{equation}

 \noindent where $\vec{b}(t)$ is a two-dimensional Brownian motion with an autocorrelation function of 

 \begin{equation}
 R_{b}(t, s) = E\{\vec{b}(s) \cdot \vec{b}(t)\} = \min(t, s).
 \label{cor}
 \end{equation}

 \noindent 
 Comparing the phase noise of \eqn{phiint} and \eqn{phiNor}, the normalized nonlinear phase noise of \eqn{phiNor} is scaled by $\phi =  L \sigma_s^2 \phi_{\mathrm{NL}}/\kappa$, $t = z/L$ is the normalized distance, $b(t) = S(tL)/\sigma_s/\sqrt{L}$ is the normalized amplifier noise, $\vec{\xi}_0 = \vec{E}_0/\sigma_s/\sqrt{L}$ is the normalized transmitted vector, and the normalized electric field of $\vec{e}_N$ is scaled by the inverse of the noise variance.
 The SNR of the signal is $\rho_s = |\vec{\xi}_0|^2 = |\vec{E_0}|^2/(L\sigma_s^2) = |\vec{E_0}|^2/(N\sigma^2_{\mathrm{ASE}})$.

 In \eqn{phiNor}, the normalized electric field $\vec{e}_N$ is the normalized received electric field without nonlinear phase noise. 
 The actual normalized received electric field, corresponding to \fig{figyinyang}, is $\vec{e}_r = \vec{e}_N \exp( - j \phi)$.
 The actual normalized received electric field has the same intensity as that of the normalized electric field $\vec{e}_N$, i.e., $|\vec{e}_r|^2 = |\vec{e}_N|^2$.
  The values of $y = |\vec{e}_N|^2$ and $r = |\vec{e}_N|$ are called normalized received intensity and amplitude, respectively. 

  \subsection{Series Expansion}

  The Brownian motion of $\vec{b}(t)$ can be expanded using the standard Karhunen-Lo\'{e}ve expansion of \cite[Sec. 10-6]{papoulis2}

  \begin{equation}
  \vec{b}(t) = \sum_{k = 1}^{\infty} \sigma_k \vec{x}_k \psi_k(t),
  \label{sum}
  \end{equation} 

  \noindent where $\vec{x}_k$ are i.i.d.~two-dimensional circular Gaussian random variables with zero mean and unity variance of $E\{|\vec{x}_k|^2\} = 1$, the eigenvalues and eigenfunctions of $\sigma_k^2, \psi_k(t),  0 \leq t \leq 1$ are \cite{ho0308} \cite[p. 305]{papoulis2} 

  \begin{equation}
  \sigma_k = \frac{2}{(2k-1) \pi},  
  \psi_k(t) = \sqrt{2} \sin \left[ \frac{(2 k - 1) \pi}{2} t \right].
  \label{sigmak}
  \end{equation}

  Substitute \eqn{sum} with \eqn{sigmak} into the normalized phase of \eqn{phiNor}, because $\int_{0}^{1} \sin(t/\sigma_k) \ud t = \sigma_k$, we get

  \begin{equation}
  \phi = |\vec{\xi}_0|^2 +  2 \sqrt{2} \sum_{k = 1}^{\infty} \sigma_k^2 \vec{\xi}_0 \cdot \vec{x}_k +  \sum_{k = 1}^{\infty} \sigma_k^2 |\vec{x}_k|^2.
  \label{sum_x}
  \end{equation}

  \noindent Because $\sum_{k = 1}^{\infty} \sigma_k^2 = \frac{1}{2}$ [see \cite[Sec. 0.234]{table}], we get

  \begin{equation}
  \phi = \sum_{k = 1}^{\infty} \sigma_k^2 | \sqrt{2} \vec{\xi}_0 +  \vec{x}_k|^2.
  \label{sum_x1}
  \end{equation}

  The random variable $|\sqrt{2} \vec{\xi}_0 +  \vec{x}_k|^2$ is a noncentral chi-square ($\chi^2$) random variable with two degrees of freedom with a noncentrality parameter of $2 \rho_s$ and a variance parameter of $\frac{1}{2}$ \cite[pp. 41-46]{proakis4}. 
  The normalized nonlinear phase noise is the summation of infinitely many independently distributed noncentral $\chi^2$-random variables with two degrees of freedom with noncentrality parameters of $ 2 \sigma_k^2 \rho_s$ and variance parameters of $\frac{1}{2} \sigma_k^2$.
  The mean and standard deviation (STD) of the random variables are both proportional to the square of the reciprocal of all odd natural numbers. 

  Using the series expansion of \eqn{sum}, the normalized electric field is

  \begin{equation}
  \vec{e}_N = \vec{\xi}_0 + \sqrt{2} \sum_{k=1}^{\infty} (-1)^{k+1} \sigma_k \vec{x}_k. 
  \end{equation}

  \noindent Using \cite[Sec. 0.232]{table}, we get $\sum_{k=1}^{\infty} (-1)^{k+1} \sigma_k = \frac{1}{2}$ and

  \begin{equation}
  \vec{e}_N = \sqrt{2} \sum_{k=1}^{\infty} (-1)^{k+1} \sigma_k (\sqrt{2} \vec{\xi}_0 + \vec{x}_k).
  \label{sum_x2}
  \end{equation}

  \noindent The normalized electric field of \eqn{sum_x2} is a two-dimensional Gaussian-distributed random variable having a mean of of $\vec{\xi}_0$ and variance of $\frac{1}{2}$.

  The series expansion of \eqn{sum_x1} and \eqn{sum_x2} can be used to derived the joint characteristic function of the normalized electric field of $\vec{e}_N$ and nonlinear phase noise of $\phi$ in \eqn{phiNor}.

  \subsection{Joint Characteristic Function}

  The joint characteristic function of the normalized nonlinear phase noise and the electric field of \eqn{phiNor} is 

  \begin{equation}
  \Psi_{\Phi, \vec{E}_N}(\nu, \vec{\omega})  =  
	  E\left\{\exp\left( j \nu \phi + j \vec{\omega} \cdot\vec{e_N}\right)\right\}. 
  \end{equation}

  \noindent This joint characteristic function was derived by \cite{mecozzi94, mecozzi94a} based on the method of \cite{cameron45}. 
  For completeness, a brief derivation is provided here using a significantly different method to eliminate some minor errors in \cite{mecozzi94, mecozzi94a, cameron45}.

  First of all, we have

  \begin{eqnarray}
  \lefteqn{E\left\{\exp\left[ j \nu   | \sqrt{2} \vec{\xi}_0 +  \vec{x}_k|^2 
	  + j  \vec{\omega} \cdot (\sqrt{2} \vec{\xi}_0 + \vec{x}_k ) \right] \right\} }\nonumber \\
  && = \frac{1}{1 - j \nu}  \exp \left( \frac{ 2 j \nu | \vec{\xi}_0|^2 + \sqrt{2} j \vec{\xi}_0 \cdot \vec{\omega} -\frac{1}{4} | \vec{\omega} |^2} {1 - j \nu} \right). \nonumber 
  \end{eqnarray}

  \noindent In the above expression, if $\vec{\omega} = 0$, the characteristic function of  $| \sqrt{2} \vec{\xi}_0 +  \vec{x}_k|^2$ is

  \begin{equation}
  \Psi_{| \sqrt{2} \vec{\xi}_0 +  x_k|^2}(\nu) = \frac{1}{1 - j \nu} 
	  \exp \left( \frac{ 2 j \nu \rho_s} {1 - j \nu} \right)
  \end{equation}

  \noindent for a noncentral $\chi^2$-distribution with mean and variance of $2 \rho_s + 1$ and $4 \rho_s + 1$, respectively \cite[p. 42]{proakis4}.

  The joint characteristic function of $\Psi_{\Phi, \vec{E}_N}$ is

  \begin{eqnarray}
  \lefteqn {\Psi_{\Phi, \vec{E}_N}(\nu, \vec{\omega})  = \prod_{k=1}^{\infty}
	  \frac{1}{1 - j \nu \sigma_k^2}  }\nonumber \\
  &&\times
	  \exp \left[ \frac{ 2 j \nu |\vec{\xi}_0|^2 \sigma_k^2 + 2j (-1)^{k+1} \sigma_k \vec{\xi}_0 \cdot \vec{\omega} - \frac{1}{2} \sigma_k^2 |\vec{\omega}|^2 } {1 - j \nu \sigma_k^2} \right] \nonumber \\
  \label{eig_phi}
  \end{eqnarray}

  \noindent as the product of the joint characteristic function of the corresponding independently distributed random variables in the series expansion of \eqn{sum_x1} and \eqn{sum_x2}.

  Using the expressions of \cite[Secs. 1.431, 1.421, 1.422]{table} 

  \begin{eqnarray}
  \cos x & = & \prod_{k=1}^{\infty} \left( 1 - \frac{4 x^2}{(2 k - 1)^2 \pi^2} \right), \nonumber \\
  \tan \frac{\pi x}{2} & = & \frac{4 x}{\pi} \sum_{k=1}^{\infty} \frac{1}{(2 k -1)^2 - x^2}, \nonumber \\  
  \sec \frac{\pi x}{2} & = & \frac{4}{\pi} \sum_{k=1}^{\infty} \frac{(-1)^{k+1} (2k-1)}{(2 k -1)^2 - x^2}, \nonumber
  \end{eqnarray}

  \noindent the characteristic function of \eqn{eig_phi} can be simplified to

  \begin{eqnarray}
  \lefteqn{ \Psi_{\Phi, \vec{E}_N}(\nu, \vec{\omega})  = \sec\!\sqrt{j \nu}} \nonumber \\
  && \exp\Bigg[\left( |\vec{\xi}_0|^2  \sqrt{j \nu} - \frac{|\vec{\omega}|^2}{4\sqrt{j \nu}}  \right) \tan\!\sqrt{j \nu}   \nonumber \\
 & & \qquad \qquad + j \sec(\sqrt{j \nu})  \vec{\xi}_0 \cdot \vec{\omega} 
	  \Bigg]. 
 \label{cfPhiEN}
 \end{eqnarray}

 \noindent The trigonometric function with complex argument is calculated by, for example, 

\[
\left( \sec \sqrt{j \nu} \right)^{-1} = \cos\sqrt{\frac{\nu}{2}} \cosh\sqrt{\frac{\nu}{2}} - j \sin\sqrt{\frac{\nu}{2}} \sinh\sqrt{\frac{\nu}{2}}.
\]

 The p.d.f.~of $p_{\Phi, \vec{E}_N}(\phi, \vec{z})$ is the inverse Fourier transform of the characteristic function $ \Psi_{\Phi, \vec{E}_N}(\nu, \vec{\omega})$ of \eqn{cfPhiEN}. 
 So far, there is no analytical expression for the p.d.f.~of $p_{\Phi, \vec{E}_N}(\phi, \vec{z})$.

 It is also obvious that  

 \begin{equation}
 \Psi_{\vec{E}_N}(\vec{\omega})  = \Psi_{\Phi, \vec{E}_N}(0, \vec{\omega}) = 
	 \exp\left( j \vec{\xi}_0 \cdot \vec{\omega} - \frac{|\vec{\omega}|^2}{4} \right)
 \end{equation}

 \noindent  is the characteristic function of a two-dimensional Gaussian distribution \cite[pp. 48-51]{proakis4} for the normalized electric field of \eqn{sum_x2}

 In the field of lightwave communications, the approach here to derive the joint characteristic function of normalized nonlinear phase noise and electric field is similar to that of \cite{foschini91} to find the joint characteristic function for polarization-mode dispersion \cite{poole91}, or that of \cite{foschini88} for filtered phase noise.
 Another approach is to solve the Fokker-Planck equation of the corresponding diffusion process \cite{gardiner2}.

 \section{The Probability Density of Nonlinear Phase Noise}
 \label{sec:non}

 The characteristic function of the normalized nonlinear phase noise is $\Psi_{\Phi, \vec{E}_N}(\nu, 0)$ or \cite{ho0308}

 \begin{equation}
 \Psi_\Phi(\nu) = \sec\!\sqrt{j \nu} \exp \left[\rho_s \sqrt{j \nu} \tan\!\sqrt{j \nu}  \right].
 \label{cfPhi}
 \end{equation}

 \noindent From the characteristic function of \eqn{cfPhi}, the mean normalized nonlinear phase shift is

 \begin{equation}
 <\!\!\Phi\!\!> = - j \left. \frac{ d}{d \nu}  \Psi_\Phi(\nu) \right|_{\nu = 0} =\rho_s + \frac{1}{2}.
\label{meanPhi}
 \end{equation}

 \noindent  Note that the differentiation or partial differentiation operation can be handled by most symbolic mathematical software. 
 The scaling from normalized nonlinear phase noise to the nonlinear phase noise of \eqn{phiint} is

 \begin{equation}
 \phi_{\mathrm{NL}} = \frac{<\!\!\Phi_{\mathrm{NL}}\!\!>}{ \rho_s + \frac{1}{2}} \phi.
 \label{phiscale}
 \end{equation}

 \noindent
 The second moment of the nonlinear phase noise is

 \begin{equation}
 <\!\!\Phi^2\!\!> = -  \left. \frac{ d^2}{d \nu^2}  \Psi_\Phi(\nu) \right|_{\nu = 0} =\frac{2}{3} \rho_s  + \frac{1}{6} +\left(\rho_s  + \frac{1}{2} \right)^2,
 \end{equation}

 \noindent that gives the variance of normalized phase noise as 

 \begin{equation}
 \sigma_\Phi^2 = \frac{2}{3}\rho_s + \frac{1}{6}.
 \label{sigmaPhi}
 \end{equation}

 \begin{figure}
 \centerline{\includegraphics[width = 0.4 \textwidth]{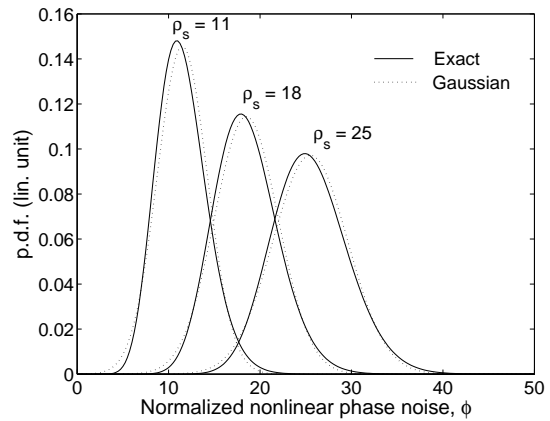}}
 \caption{The p.d.f.~of the normalized nonlinear phase noise $\phi$ for SNR of $\rho_s = 11, 18$, and $25$.}
 \label{pdfsnr}
 \end{figure}

 The p.d.f.~of the normalized nonlinear phase noise of \eqn{phiNor} can be calculated by taking the inverse Fourier transform of the characteristic function \eqn{cfPhi}. 
 \fig{pdfsnr} shows the p.d.f.~of the normalized nonlinear phase noise for three different SNR of $\rho_s = 11, 18,$ and $25$ (10.4, 12.6 and 14.0 dB), corresponding to about an error probability of $10^{-6}$, $10^{-9}$, and $10^{-12}$, respectively, when amplifier noise is the sole impairment. 
 \fig{pdfsnr} shows the p.d.f.~using the exact characteristic function \eqn{cfPhi}, and the Gaussian approximation with mean and variance of $<\!\!\Phi\!\!>$ \eqn{meanPhi} and $\sigma_\Phi^2$ \eqn{sigmaPhi}.  
 From \fig{pdfsnr}, the Gaussian distribution is not a good model for nonlinear phase noise.

 \begin{figure}
 \centerline{\includegraphics[width = 0.4 \textwidth]{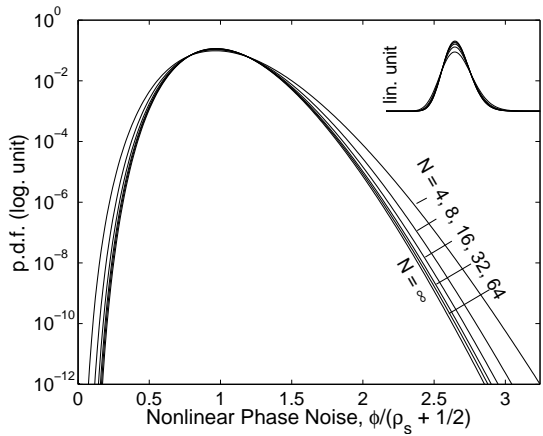}}
 \caption{The asymptotic p.d.f.~of normalized nonlinear phase noise of $\phi$ as compared with the p.d.f.~of $N=4, 8, 16, 32,$ and $64$ fiber spans.
 The p.d.f.~in linear scale is shown in the inset.}
 \label{pdfN}
 \end{figure}

 The p.d.f.~for finite number of fiber spans was derived base on the orthogonalization of the nonlinear phase noise of \eqn{phiNL} by the summation of $N$ independently distributed random variables \cite{ho0309c}.  
 \fig{pdfN} shows a comparison of the p.d.f.~for  $N = 4, 8, 16, 32$, and $64$ of fiber spans \cite{ho0309c} with the asymptotic case of \eqn{cfPhi}. 
 Using the SNR of $\rho_s = 18$ (12.6 dB), \fig{pdfN} is plotted in logarithmic scale to show the difference in the tail.
 \fig{pdfN} also provides an inset in linear scale of the same p.d.f.~to show the difference around the mean.
 The asymptotic p.d.f.~of \eqn{cfPhi} with distributed noise has the smallest spread in the tail as compared with the p.d.f.~with $N$ discrete noise sources. 
 The asymptotic p.d.f.~is very accurate for $N \geq 32$ fiber spans. 

Comparing \figs{pdfsnr} and \ref{pdfN}, when the p.d.f.~is plotted in linear scale, the difference between the actual and Gaussian approximation seems small as compared with the same p.d.f.~plots in logarithmic scale.
In linear scale, the p.d.f. also seems more symmetric in \fig{pdfsnr} or the inset of \fig{pdfN}.
When phase noise is very large or plots in linear scale \cite{holzlohner02, hanna01}, the Gaussian approximation seems more valid.

 As discussed earlier, the effects of amplifier noise outside the signal bandwidth and the amplifier noise from orthogonal polarization are all ignored for simplicity. 
 If the nonlinear phase noise induced from those amplifier noises is included, based on the simple reasoning of \cite{humblet91}, the marginal characteristic function of the normalized nonlinear phase noise of \eqn{cfPhi} becomes

 \begin{equation}
  \sec^{\frac{m}{2}}\left(\sqrt{j \nu}\right) \exp \left[\rho_s \sqrt{j \nu} \tan\!\sqrt{j \nu}  \right].
 \label{cfPhiMany}
 \end{equation}

 \noindent where $m$ is product of the ratio of the amplifier noise bandwidth to the signal bandwidth and the number of polarizations. 
 If only the amplifier noise from orthogonal polarization matched to signal bandwidth is also considered, $m =2$ for two polarizations gives the charactersitic function of \eqn{cfPhi}.
 With cross-phase modulation induced nonlinear phase noise, the mean and variance of the nonlinear phase noise increase slightly to $\rho_s  + \frac{1}{2} m$ and $ \frac{2}{3}\rho_s  + \frac{1}{6} m$, respectively.
 The nonlinear phase noise is induced mainly by the beating of the signal and amplifier noise from the same polarization as the signal, similar to the case of signal-spontaneous beat noise in an amplified receiver.
 For high SNR of $\rho_s$, it is obvious that the signal-amplifier noise beating is the major contribution to nonlinear phase noise.
 The parameter of $m$ can equal to $1$ for the case if the amplifier noise from another dimension is ignored by confining to single-dimensional signal and noise.
 In later part of this article, the characteristic function of \eqn{cfPhi} can be changed to \eqn{cfPhiMany} if necessary.

 The characteristic function of \eqn{cfPhiMany} assumes a dispersionless fiber.
 With fiber dispersion, due to walk-off effect, the nonlinear phase noise caused by cross-phase modulation should approximately have Gaussian distribution. 
 Method similar to \cite{chiang96, ho0007} can be used to find the variance of the nonlinear phase noise due to cross-phase modulation in dispersive fiber.
 For either PSK and DPSK signals, the signal induced nonlinear phase noise by cross-phase modulation should be very small. 
 The power spectral density of signal and noise can be first derived, multiplied by the transfer function due to walk-off from \cite{chiang96, ho0007} and integrated over all frequency gives the variance of phase noise.

For DPSK signal, the phase noises in adjacent symbols are correlated to each other \cite{mckinstrie03}.  
The characteristic function of the differential phase due to cross-phase modulation can be found using the power spectral density of \cite{chiang96}, taking the inverse Fourier transform to get the autocorrelation function, and getting the correlation coefficient as the autocorrelation with a time difference of the symbol interval. 
The characteristic function of the differential phase decreases by the correlation coefficient.

Similar to \cite{gordon90, ho0309c, ho0308}, all the derivations here assume non-return-to-zero (NRZ) pulses (or continuous-wave signal) but most experiments \cite{gnauck02, zhu02, miyamoto02, bissessur03, gnauck03, cho03, rasmussen03, zhu03, vareille03, tsuritani03, cai03} use return-to-zero (RZ) pulses.
For flat-top RZ pulse, the mean nonlinear phase shift of $<\!\!\Phi_{\mathrm{NL}}\!\!>$ should be the mean nonlinear phase shift when the peak amplitude is transmitted.  
Usually, $<\!\!\Phi_{\mathrm{NL}}\!\!>$ is increased by the inverse of the duty cycle.
However, for soliton and dispersion-managed soliton , based on soliton perturbation  \cite{kivshar89, kaup90, georges95, iannone} or variational principle \cite{mckinstrie02, mckinstrie02a}, the mean nonlinear phase shift of $<\!\!\Phi_{\mathrm{NL}}\!\!>$ is reduced by a factor of 2 when dispersion and self-phase modulation balance each other \cite{ho0403b}.

 \section{Some Joint Characteristic Functions}
 \label{sec:jcf}

 From the characteristic function \eqn{cfPhiEN}, we can take the inverse Fourier transform with respect to $\vec{\omega}$ and get

 \begin{equation}
 \mathcal{F}^{-1}_{\vec{\omega}} \left\{ \Psi_{\Phi, \vec{E}_N}(\nu, \vec{\omega}) \right\}
  = \mathcal{F}_{\phi} \left\{ p_{\Phi, \vec{E}_N}(\phi, \vec{z}) \right\}, 
\end{equation}

 \noindent where $\mathcal{F}^{-1}_{\vec{\omega}} $ denotes the inverse Fourier transform with respect to $\vec{\omega}$, and $\mathcal{F}_{\phi}$ denotes the Fourier transform with respect to $\phi$.
 The characteristic function of \eqn{cfPhiEN} can be rewritten as

\begin{eqnarray}
 \Psi_{\Phi, \vec{E}_N}(\nu, \vec{\omega})  = \Psi_\Phi(\nu)  \exp\Bigg[ -\frac{|\vec{\omega}|^2 \tan\!\sqrt{j \nu}}{4\sqrt{j \nu}}  \nonumber \\
+ j \sec(\sqrt{j \nu}) \vec{\xi}_0 \cdot \vec{\omega} 
	 \Bigg]. 
\label{cfsimp1}
\end{eqnarray}

\noindent where $\Psi_\Phi(\nu)$ is the marginal characteristic function of nonlinear phase noise from \eqn{cfPhi}. 
The inverse Fourier transform is

\begin{eqnarray}
\mathcal{F}^{-1}_{\vec{\omega}} \left\{ \Psi_{\Phi, \vec{E}_N} \right\}   =   \frac{\Psi_\Phi(\nu)}{ 2 \pi \sigma_\nu^2} \exp\left( -\frac{|\vec{z} -  \vec{\xi}_\nu|^2}{2 \sigma_\nu^2} \right), 
\label{cfPhipdfz}
\end{eqnarray}

\noindent where $\sigma_\nu^2 = \frac{1}{2} \tan( \sqrt{j \nu} )/\sqrt{j \nu}$ and $\vec{\xi}_\nu = \sec(\sqrt{j \nu}) \vec{\xi}_0$. 
Both $\sigma_\nu^2$ and $\vec{\xi}_\nu$ are complex numbers and can be considered as the angular frequency depending variance and mean, respectively.

\subsection{Joint Characteristic Function of Nonlinear Phase Noise and Received Intensity}

Using the partial p.d.f.~and characteristic function of \eqn{cfPhipdfz}, change the random variable from rectangular coordinate of $\vec{z} = (z_1, z_2)$ to polar coordinate of $\vec{z} = (r\cos\theta, r\sin\theta)$, we get

\begin{eqnarray}
\lefteqn{ \mathcal{F}^{-1}_{\vec{\omega}} \left\{ \Psi_{\Phi, R, \Theta_n} \right\}   
	=   \frac{r \Psi_\Phi(\nu)}{ 2 \pi \sigma_\nu^2} } \nonumber \\
&& \times \exp\left[ -\frac{r^2 + | \vec{\xi}_\nu|^2 - 2 r | \vec{\xi}_\nu| \cos(\theta - \theta_0) }{2 \sigma_\nu^2} \right], 
\label{cfPhipdfpolar}
\end{eqnarray}

\noindent where $\theta_0$ is the angle of the transmitted vector $\vec{\xi}_0$ and $| \vec{\xi}_\nu| = \sec(\sqrt{j \nu}) |\vec{\xi}_0|$.
The random variable of $\Theta_n$ is called the phase of amplifier noise because it is solely contributed from amplifier noise.

Taking the integration over $\theta$ and changing the random variable to the received intensity of $y = r^2$, we get

\begin{eqnarray}
 \mathcal{F}^{-1}_{\omega} \left\{ \Psi_{\Phi, Y} \right\}   =   \frac{\Psi_\Phi(\nu)}{ 2 \sigma_\nu^2} \exp\!\!\left[ -\frac{y + | \vec{\xi}_\nu|^2  }{2 \sigma_\nu^2} \right] 
I_0\!\! \left[ \sqrt{y} \frac{|\vec{\xi}_\nu|} { \sigma_\nu^2} \right], \nonumber \\
\label{cfPhipdfY}
\end{eqnarray}

\noindent where $I_k(\cdot)$ is the $k^{\mathrm{th}}$-order modified Bessel function of the first kind. 
The p.d.f.~of the received intensity of

\begin{eqnarray}
p_{Y}(y)  & = &  \left. \mathcal{F}^{-1}_{\omega} \left\{ \Psi_{\Phi, Y} \right\}(\nu, y) \right|_{\nu = 0} \nonumber \\
	& = & \exp\left( - y - \rho_s \right) I_0 \left( 2 \sqrt{y \rho_s} \right)
\label{pdfY}
\end{eqnarray}	

\noindent is a non-central $\chi^2$-p.d.f.~with two degrees of freedom with a noncentrality parameter of $\rho_s$ and variance parameter of $\frac{1}{2}$ \cite[pp. 41-44]{proakis4}. 
With a change of random variable of $y = r^2$, the received amplitude has a Rice distribution of \cite[pp. 46-47]{proakis4}

\begin{equation}
p_R(r) = 2 r \exp\left[-(r^2 + \rho_s)\right] I_0(2 r \sqrt{\rho_s}).
\label{pdfR}
\end{equation}

Taking a Fourier transform of \eqn{cfPhipdfY}, the joint characteristic function of nonlinear phase noise and received intensity is

\begin{eqnarray}
\Psi_{\Phi, Y}(\nu, \omega)  =   \frac{\Psi_\Phi(\nu)}{ 1 -2 j  \omega \sigma_\nu^2} 
\exp \left[ \frac{j \omega |\vec{\xi}_\nu|^2} {1 - 2 j \omega \sigma_\nu^2}  \right],
\end{eqnarray}

\noindent or

\begin{eqnarray}
\lefteqn{ \Psi_{\Phi, Y}(\nu, \omega)  = \frac{1}
	{\cos\!\sqrt{j \nu} - j \omega \frac{\sin\!\sqrt{j \nu}}{ \sqrt{j \nu} }} } \nonumber \\
&& \quad \times \exp \Bigg[\rho_s  \sqrt{j \nu} \tan\!\sqrt{j \nu} \nonumber \\
&& \qquad + \frac{ j \omega \rho_s }
	{\cos^2\!\sqrt{j \nu} - j \omega \frac{ \sin(2\sqrt{ j \nu})} {2\sqrt{j \nu}}}
	\Bigg].	
\label{cfPhiY}
\end{eqnarray}

\noindent The joint characteristic function of \eqn{cfPhiY} can be used to study the compensation of nonlinear phase noise using received intensity \cite{ho0403a, liu02, xu02}. 

\subsection{Joint Characteristic Function of Nonlinear Phase Noise and the Phase of Amplifier Noise}
\label{sec:cfPhiYTheta}

Using the characteristic function of \eqn{cfPhipdfpolar}, take the integration over the received amplitude $r$, we get 

\begin{eqnarray}
\lefteqn{ \mathcal{F}^{-1}_{\omega} \left\{ \Psi_{\Phi, \Theta_n} \right\}  
	=   \frac{\Psi_\Phi(\nu)}{ 2 \pi \sigma_\nu^2} } \nonumber \\
&& \times \int_0^\infty \exp\left[ -\frac{r^2 + | \vec{\xi}_\nu|^2 - 2 r | \vec{\xi}_\nu| \cos(\theta - \theta_0) }{2 \sigma_\nu^2} \right] r \ud r, \nonumber
\end{eqnarray}

\noindent or

\begin{eqnarray}
\lefteqn{ \mathcal{F}^{-1}_{\omega} \left\{ \Psi_{\Phi, \Theta_n} \right\}   =   \Psi_\Phi(\nu) }\nonumber \\
&& \quad \times \Bigg\{ \frac{1}{2 \pi}e^{-\gamma_\nu}   +  \sqrt{\frac{\gamma_\nu}{4 \pi}} \cos(\theta - \theta_0)  e ^{-\gamma_\nu \sin^2(\theta - \theta_0)}  \nonumber \\
&& \qquad \qquad \times \mathrm{erfc} \left[ - \sqrt{\gamma_\nu} \cos(\theta - \theta_0) \right] \Bigg\},
\label{cfPhipdfTheta}
\end{eqnarray}

\noindent where 

\begin{equation}
\gamma_\nu = \frac{| \vec{\xi}_\nu|^2}{2\sigma_\nu^2} = \frac{2 \sqrt{j \nu}} {\sin\left( 2 \sqrt{j \nu}\right)} \rho_s
\label{gammanu}
\end{equation}

\noindent can be interpreted as the angular frequency depending SNR. 

Taking the Fourier transform of \eqn{cfPhipdfTheta}, from \cite[Sec. 9.2-2]{middleton}, the characteristic function of $\Psi_{\Phi, \Theta_n}$ is

\begin{eqnarray}
\lefteqn{ \Psi_{\Phi, \Theta_n}(\nu, \omega) = \Psi_\Phi(\nu)}  \nonumber \\
& & \times  \sum_{m = 0}^{\infty} \epsilon_m \frac{\gamma_\nu^{\frac{m}{2}}}{2 m!} \Gamma\left(\frac{m}{2} + 1\right)
	{}_1F_1\left( \frac{m}{2};m+1; -\gamma_\nu \right) \nonumber \\
&& \quad \times \bigg[ \frac{(e^{2 \pi j (m + \omega)} - 1) e^{-j m \theta_0}}{2 j\pi (m + \omega)} \nonumber \\
& & \qquad \qquad	+ \frac{(e^{2 \pi j (\omega-m)} - 1) e^{j m \theta_0}}{2 j \pi (\omega-m)}
	\bigg] . 
\label{cfPhiTheta}
\end{eqnarray}

\noindent where $\Gamma(\cdot)$ is the Gamma function, ${}_1F_1(a; b; \cdot)$ is the confluent hypergeometric function of the first kind with parameters of $a$ and $b$, and $\epsilon_m = 1$ if $m = 1$,  $\epsilon_m =2$ if $m \geq 1$.

Within the summation of the joint characteristic function of \eqn{cfPhiTheta}, if $\omega = m$ as an integer, only one term in the summation is non-zero.
We have 

\begin{eqnarray}
\lefteqn {\Psi_{\Phi, \Theta_n}(\nu, m)  =   
	\frac{\Psi_\Phi(\nu) \gamma_\nu^{\frac{m}{2}}}{m!} \Gamma\!\!\left(\frac{m}{2} + 1\right) }\nonumber \\
	& &	\qquad \times 	{}_1F_1\!\!\left( \frac{m}{2};m+1; -\gamma_\nu \right)e^{j m \theta_0} \nonumber \\
& = & \frac{\sqrt{\pi}}{2} \Psi_\Phi(\nu)  \gamma_\nu^{\frac{1}{2}} \exp\left(-\frac{\gamma_\nu}{2} \right) \nonumber \\
& & \qquad \times \left[I_{\frac{m-1}{2}}\left( \frac{\gamma_{\nu}}{2} \right) 
    + I_{\frac{m+1}{2}}\left( \frac{\gamma_{\nu}}{2} \right) \right] e^{j m \theta_0}, \nonumber \\
& & \quad m \geq 0,
\label{cfPhiThetam}
\end{eqnarray}

\noindent and $\Psi_{\Phi, \Theta_n}(\nu, -m) = \Psi_{\Phi, \Theta_n}(\nu, m)e^{-2 j m \theta_0}$.
Using \cite[Sec. 9.212, Sec. 9.238]{table}, the conversion from hypergeometric function to Bessel functions in \eqn{cfPhiThetam} is used in \cite{jain73, jain74, blachman81, blachman88}. 
The simple expression for $\Psi_{\Phi, \Theta_n}(\nu, m)$ is very helpful to derive the p.d.f.~of the signal phase.
The coefficients of \eqn{cfPhiThetam} are also the Fourier series coefficients of the expression of \eqn{cfPhipdfTheta} expanded over the phase $\theta$ in the range of $[-\pi, \pi)$.

\subsection{Joint Characteristic Functions of Nonlinear Phase Noise, Received Intensity and the Phase of Amplifier Noise}

Here, we derive the joint characteristic functions of $ \Psi_{\Phi, Y, \Theta_n}$.
Similar to \eqn{cfPhiThetam}, corresponding to the Fourier coefficients, only the characteristic function at integer ``angular frequency'' of $\Theta_n$ is interested. 
With $m$ as an non-negative integer, using  \cite[Sec. 8.431]{table} and \eqn{cfPhipdfpolar} with $y = r^2$, we get

\begin{eqnarray}
\lefteqn{ \mathcal{F}^{-1}_{\omega} \left\{ \Psi_{\Phi, Y, \Theta_n}(\nu, \omega, m) \right\}
   	=   \frac{\Psi_\Phi(\nu) e^{j m \theta_0} }{ 2  \sigma_\nu^2}  }\nonumber \\
&& \quad \times  
        \exp\left[ -\frac{y + | \vec{\xi}_\nu|^2  }{2 \sigma_\nu^2} \right] 
        I_m \left( \frac{\sqrt{y} | \vec{\xi}_\nu|}{\sigma_\nu^2} \right), m \geq 0,
\label{cfPhiThetampdfY}
\end{eqnarray}

\noindent Taking the Fourier transform of \eqn{cfPhiThetampdfY}, we get

\begin{eqnarray}
\lefteqn{ \Psi_{\Phi, Y, \Theta_n}(\nu, \omega, m)
   	=   \frac{\Psi_\Phi(\nu) e^{j m \theta_0} }{ 2  \sigma_\nu^2}  }\nonumber \\
&& \times \int_0^\infty 
        \exp\left[ -\frac{y + | \vec{\xi}_\nu|^2  }{2 \sigma_\nu^2} \right] 
        I_m \left( \frac{\sqrt{y} | \vec{\xi}_\nu|}{\sigma_\nu^2} \right)  e^{j \omega y} \ud y, \qquad \nonumber \\
& & \qquad \qquad \qquad \qquad \qquad m \geq 0, 
\nonumber
\end{eqnarray}

\noindent and $\Psi_{\Phi, Y, \Theta_n}(\nu, \omega, -m) = \Psi_{\Phi, Y, \Theta_n}(\nu, \omega, m) e^{-2 j m \theta_0}$. 

Using \cite[Sec. 6.614, Sec. 9.220]{table}, we get

\begin{eqnarray}
\lefteqn{ \Psi_{\Phi, Y, \Theta_n}(\nu, \omega, m)   =    \Psi_\Phi(\nu)
\exp\left[ -\frac{| \vec{\xi}_\nu|^2  }{2 \sigma_\nu^2} \right]
\frac{\Gamma(\frac{1}{2}m + 1)}{m!}  } \nonumber \\
& & \times  \frac{\sqrt{2} \sigma_\nu  e^{j m \theta_0} \gamma_{\nu, \omega}^{\frac{m}{2} + \frac{1}{2}} } { (1 - 2j \omega \sigma_\nu^2)^{\frac{1}{2}}| \vec{\xi}_\nu|  } 
    {}_1F_1\left( \frac{m}{2}+1; m+1; \gamma_{\nu, \omega} \right), \nonumber \\
& & \qquad \qquad \qquad m \geq 0,
\label{cfPhiYThetam1}
\end{eqnarray}

\noindent where 

\begin{eqnarray}
\gamma_{\nu, \omega} &=& \frac{1}{1 - 2j \omega \sigma_\nu^2} \frac{ |\vec{\xi}_\nu|^2 } {2 \sigma_\nu^2} \nonumber \\
   & =  &  \frac{2 j \nu} {\left[\sqrt{j \nu} - j \omega  \tan( \sqrt{j \nu}) \right] \sin\left( 2 \sqrt{j \nu}\right)} \rho_s. 
\label{gammano}
\end{eqnarray}

Using $\gamma_\nu$ defined by \eqn{gammanu}, we can rewite \eqn{cfPhiYThetam1} as

\begin{eqnarray}
\lefteqn {\Psi_{\Phi, Y, \Theta_n}(\nu, \omega, m)   =    \Psi_\Phi(\nu) 
\frac{e ^{-\gamma_\nu+j m \theta_0} \Gamma(\frac{1}{2}m + 1)}{ m!} } \nonumber \\
& & \qquad \times \frac{\gamma_{\nu, \omega}^{\frac{m}{2}+1}}  {\gamma_\nu}  
{}_1F_1\left( \frac{m}{2}+1; m+1; \gamma_{\nu, \omega} \right). 
\label{cfPhiYThetam2}
\end{eqnarray}

With $\omega = 0$ in \eqn{cfPhiYThetam2}, using  \cite[Sec. 9.212]{table}, the joint Fourier coefficients of nonlinear phase noise and the phase of amplifier noise are

\begin{eqnarray}
\lefteqn {\Psi_{\Phi, \Theta_n}(\nu, m)   =    \Psi_\Phi(\nu) 
\frac{e ^{-\gamma_\nu + j m \theta_0} \Gamma(\frac{1}{2}m + 1)}{ m! } \gamma_\nu^{\frac{m}{2}} }\nonumber \\
& & \qquad \times {}_1F_1\left( \frac{m}{2}+1; m+1; \gamma_\nu \right) \nonumber \\
& = & \frac{\Psi_\Phi(\nu) \gamma_{\nu}^{\frac{m}{2}}  e^{j m \theta_0}}{m!} \Gamma\!\!\left(\frac{m}{2} + 1\right){}_1F_1\!\!\left( \frac{m}{2};m+1; -\gamma_{\nu} \right), \nonumber 
\end{eqnarray}

\noindent the same as \eqn{cfPhiThetam}.

If $m = 0$ in \eqn{cfPhiYThetam2},  using the relationship of ${}_1 F_1(1; 1; z) = e^z$ \cite[Sec. 9.215]{table}, the joint characteristic function of nonlinear phase noise and the received intensity is

\begin{eqnarray}
\Psi_{\Phi, Y}(\nu, \omega)  & = &   \Psi_\Phi(\nu) 
e ^{-\gamma_\nu} \frac{\gamma_{\nu, \omega}}  {\gamma_\nu} e^{\gamma_{\nu, \omega}} \nonumber \\
& = & \frac{\Psi_\Phi(\nu)}{ 1 -2 j  \omega \sigma_\nu^2} 
\exp \left[ \frac{j \omega |\vec{\xi}_\nu|^2} {1 - 2 j \omega \sigma_\nu^2}  \right],
\label{cfPhiY1}
\end{eqnarray}

\noindent also the same as \eqn{cfPhiY}.

To simplify \eqn{cfPhiYThetam2} using the Bessel functions, from \cite[Sec. 9.212, Sec. 9.238]{table} and similar to \cite{jain73, jain74, blachman81, blachman88}, we get

\begin{eqnarray}
\lefteqn {\Psi_{\Phi, Y, \Theta_n}(\nu, \omega, m)   =   \Psi_\Phi(\nu)
\frac{\sqrt{\pi} \gamma_{\nu, \omega}^{\frac{3}{2}}}{ 2 \gamma_{\nu} } 
\exp\left(-\gamma_{\nu} + \frac{\gamma_{\nu, \omega}}{2}\right) 
} 
\nonumber \\
& & \qquad \times \left[I_{\frac{m-1}{2}}\left( \frac{\gamma_{\nu, \omega}}{2} \right) 
    + I_{\frac{m+1}{2}}\left( \frac{\gamma_{\nu, \omega}}{2} \right) \right] e^{j m \theta_0}. \nonumber \\
\label{cfPhiYThetam}
\end{eqnarray}

Although both joint characteristic functions \eqn{cfPhiY} and \eqn{cfPhiThetam} can be derived based on the joint characteristic function \eqn{cfPhiYThetam2} or \eqn{cfPhiYThetam}, they are derived seperately in early parts of this section for simplicity.

\section{Error Probability of Phase-Modulated Signals}
\label{sec:ber}

Binary DPSK signaling with interferometer based direct-detection receiver has renewed interests recently \cite{gnauck02, zhu02, miyamoto02, bissessur03, zhu03, gnauck03}.
This section studies the impact of nonlinear phase noise to binary PSK and DPSK signals.
In order to derive the error probability, the p.d.f.~of the received phase, that is the summation of both nonlinear phase noise and the phase of amplifier noise, is first derived analytically as a Fourier series. 
Taking into account the dependence between the nonlinear phase noise and the phase of amplifier noise, the error probability is calculated using the Fourier coefficients.

If the nonlinear phase noise is assumed to be Gaussian distributed, the error probability is the same as that of \cite{nicholson84} with laser phase noise.
Because laser phase noise is a Brownian motion, the phase noise difference between two consecutive symbols is Gaussian distributed.

The optimal operating point of the system is estimated by \cite{gordon90} based on the insight that the variance of linear and nonlinear phase noise should be approximately the same.
With the exact error probability, the system can be optimized rigorously by the condition that the increase in SNR penalty is less than the increase of launched power.

\subsection{Phase Distribution}

Without loss of generality, in this section, the normalized transmitted electric field is assumed to be $\vec{\xi}_0 = (\sqrt{\rho_s}, 0)$ when $\theta_0 = 0$.
With nonlinear phase noise, received by an optical phase-locked loop \cite{norimatsu92, kahn90}, the overall received phase is 
 
\begin{equation}
\Phi_r = \Theta_n - \Phi_{\mathrm{NL}} =\Theta_n - 
		\frac{ <\!\! \Phi_{\mathrm{NL}}\!\!>} {\rho_s + \frac{1}{2}} \Phi,
\label{PhiR}
\end{equation}

\noindent where $\rho_s + \frac{1}{2}$ is the mean normalized nonlinear phase shift \eqn{meanPhi}.

The received phase is confined to the range of $[-\pi, +\pi)$. 
The p.d.f. of the received phase is a periodic function with a period of $2 \pi$.
If the characteristic function of the received phase is $\Psi_{\Phi_r}(\nu)$, the p.d.f. of the received phase has a Fourier series expansion of

\begin{equation}
p_{\Phi_r}(\theta) =\frac{1}{2 \pi} \sum_{m = -\infty}^{+\infty} \Psi_{\Phi_r}(m) \exp( j m \theta).
\label{pdfPhiR1}
\end{equation}

Because the characteristic function has the property of $\Psi_{\Phi_r}(-\nu) = \Psi^*_{\Phi_r}(\nu)$, we get

\begin{equation}
p_{\Phi_r}(\theta) = \frac{1}{2 \pi} + \frac{1}{\pi} 
\sum_{m =1}^{+\infty}\Re \left\{ \Psi_{\Phi_r}(m) \exp( j m \theta) \right\},
\label{pdfPhiR}
\end{equation}

\noindent where $\Re\{\cdot\}$ denotes the real part of a complex number.

Using the joint characteristic function of $\Psi_{\Phi, \Theta_n}(\nu, m)$ \eqn{cfPhiThetam}, from the received phase of \eqn{PhiR}, the Fourier series coefficients are

\begin{equation}
 \Psi_{\Phi_r}(m) = \Psi_{\Phi, \Theta_n} \left( - 
		m \frac{ <\!\! \Phi_{\mathrm{NL}}\!\!>} {\rho_s + \frac{1}{2}}, m \right).
\label{cfPhiR}
\end{equation}

\begin{figure}
\centerline{\includegraphics[width = 0.4 \textwidth]{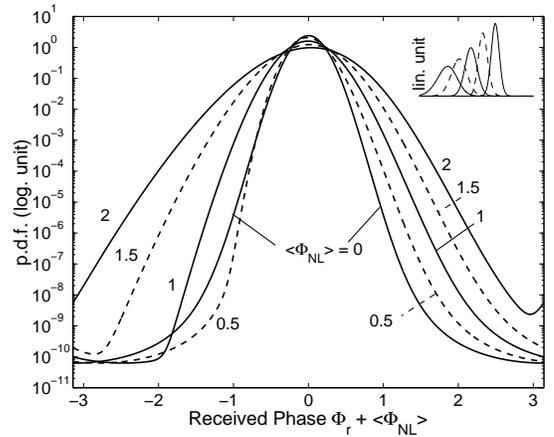}}
\caption{The p.d.f. of the received phase $p_{\Phi_r}(\theta +<\!\!\Phi_{\mathrm{NL}}\!\!>)$ in logarithmic scale. 
The inset is the p.d.f. of $p_{\Phi_r}(\theta)$ in linear scale.}
\label{figpdfPhiR}
\end{figure}

\fig{figpdfPhiR} shows the p.d.f. of the received phase \eqn{pdfPhiR} with mean nonlinear phase shift of $<\!\!\Phi_{\mathrm{NL}}\!\!> = 0, 0.5, 1.0, 1.5$, and $2.0$ rad.
Shifted by the mean nonlinear phase shift $<\!\!\Phi_{\mathrm{NL}}\!\!>$,
the p.d.f. is plotted in logarithmic scale to show the difference in the tail.
Not shifted by $<\!\!\Phi_{\mathrm{NL}}\!\!>$, the same p.d.f. is plotted in linear scale in the inset.
\fig{figpdfPhiR} is plotted for the case that the SNR is equal to $\rho_s = 18$ (12.6 dB), corresponding to an error probability of $10^{-9}$ if amplifier noise is the sole impairment.
Without nonlinear phase noise $<\!\!\Phi_{\mathrm{NL}}\!\!> = 0$, the p.d.f. is the same as that in \cite[Sec. 5.2.7]{proakis4} and symmetrical with respect to the zero phase.

From \fig{figpdfPhiR}, when the p.d.f. is broadened by the nonlinear phase noise, the broadening is not symmetrical with respect to the mean nonlinear phase shift $<\!\!\Phi_{\mathrm{NL}}\!\!>$. 
With small mean nonlinear phase shift of  $<\!\!\Phi_{\mathrm{NL}}\!\!> = 0.5$ rad, the received phase spreads further in the positive phase than the negative phase.
With large mean nonlinear phase shift of  $<\!\!\Phi_{\mathrm{NL}}\!\!> = 2$ rad, the received phase spreads further in the negative phase than the positive phase.
The difference in the spreading for small and large mean nonlinear phase shift is due to the dependence between nonlinear phase noise and the phase of amplifier noise.
As shown in \fig{pdfsnr}, after normalization, the p.d.f. of nonlinear phase noise depends solely on the SNR.
If nonlinear phase noise is independent of the phase of amplifier noise, the spreading of the received phase noise is independent of the mean nonlinear phase shift $<\!\!\Phi_{\mathrm{NL}}\!\!>$.

\subsection{Error Probability of PSK Signals}

If the p.d.f. of \eqn{pdfPhiR} were symmetrical with respect to the mean nonlinear phase shift $<\!\!\Phi_{\mathrm{NL}}\!\!>$, the decision region would center at the mean nonlinear phase shift  $<\!\!\Phi_{\mathrm{NL}}\!\!>$ and the decision angle for binary PSK system should be $\pm \frac{1}{2} \pi - <\!\!\Phi_{\mathrm{NL}}\!\!>$.
From \fig{figpdfPhiR}, because the p.d.f. is not symmetrical with respect to the mean nonlinear phase shift $<\!\!\Phi_{\mathrm{NL}}\!\!>$, assume that the decision angle is $\pm \frac{1}{2} \pi - \theta_c$ with the center phase of $\theta_c$, the error probability is

\begin{equation}
p_{e, \mathrm{PSK}} = 1 - \int_{-\frac{1}{2} \pi -\theta_c}^{ \frac{1}{2}\pi-\theta_c} p_{\Phi_r} (\theta) \ud \theta,
\end{equation} 

\noindent or

\begin{equation}
p_{e, \mathrm{PSK}} = \frac{1}{2} - \frac{1}{\pi}
\sum _{m =1}^{+\infty} \frac{2 \sin\left(\frac{1}{2}m \pi \right)}{m} \Re \left\{ \Psi_{\Phi_r}(m) e^{j m \theta_c} \right\}.
\end{equation}

After some simplifications, we get

\begin{equation}
p_{e, \mathrm{PSK}} = \frac{1}{2} - \frac{2}{\pi}
\sum _{k =0}^{+\infty}
   \frac{(-1)^k}{2 k + 1}\Re \left\{ \Psi^*_{\Phi_r}(2 k + 1) e^{ -j (2 k +1) \theta_c} \right\}.
\label{BerPSK}
\end{equation}

From the characteristic function of \eqn{cfPhiThetam}, the coefficients for the error probability \eqn{BerPSK} are

\begin{eqnarray}
\lefteqn{\Psi^*_{\Phi_r}(2k+1)  =   \frac{\sqrt{\pi \lambda_k}}{2} e^{-\frac{1}{2}\lambda_k} \left[I_k \left( \frac{\lambda_k}{2} \right) 
	+ I_{k+1} \left( \frac{\lambda_k}{2} \right) \right]} \nonumber \\
&&\qquad \qquad	\qquad \times \Psi_{\Phi} 
		\left[ \frac{(2 k + 1) <\!\!\Phi_{\mathrm{NL}}\!\!> }{ \rho_s + \frac{1}{2}} \right], \quad  k \geq 0, \qquad
\label{cfPhiRk}
\end{eqnarray}

\noindent where, from \eqn{gammanu},

\begin{eqnarray}
\lambda_k  = \frac{2 \left[ \frac{j(2 k + 1) <\!\Phi_{\mathrm{NL}}\!> }{ \rho_s + \frac{1}{2}} \right]^{\frac{1}{2}}  }
 	{\sin \left\{2 \left[\frac{j(2 k + 1) <\!\Phi_{\mathrm{NL}}\!> }{ \rho_s + \frac{1}{2}} \right]^{\frac{1}{2}} \right\} } \rho_s, k \geq 0,
\label{lambdak}
\end{eqnarray}

\noindent  are equivalent to the angular frequency depending SNR parameters.

Note that the exact error probability \eqn{BerPSK} is very similar to that in \cite{mecozzi94}. 
However,  the error probability of eq. (71) of \cite{mecozzi94} is for PSK instead of DPSK signal. This will be more clear in later parts of this paper.
The major difference between the exact error probability \eqn{BerPSK} and that in \cite{mecozzi94} is the observation that the center phase is not equal to the mean nonlinear phase shift.
From \eqn{cfPhi}, the shape of the p.d.f. of nonlinear phase noise depends solely on the signal SNR.

In the coefficients of \eqn{cfPhiRk}, the complex coefficients of $\lambda_k$ \eqn{lambdak} are equivalent to the angular frequency depending SNR parameters.
Bessel functions with complex argument are well-defined \cite{amos86}.

The coefficients of \eqn{cfPhiRk} have a very complicated expression because of the dependence between 
the phase of amplifier noise and the nonlinear phase noise.
Because $E\{\Theta_n \Phi_{\mathrm{NL}}\} = 0$, the phase of amplifier noise and the nonlinear phase noise  are uncorrelated with each other.
However, uncorrelated is not equivalent to independence for non-Gaussian random variables.

If the nonlinear phase noise is assumed to be independent to the phase of amplifier noise \cite{ho0309a}, similar to the approaches of \cite{jain74, nicholson84} in which the extra phase noise is independent of the signal phase, the error probability can be approximated as

\begin{eqnarray}
\lefteqn{ p_{e, \mathrm{PSK}} \approx \frac{1}{2} - e^{-\frac{\rho_s}{2}}  \sqrt{\frac{\rho_s}{\pi}}
	  \sum_{k = 0}^{\infty} \frac{(-1)^k} {2 k + 1}
\left[ I_{k}\!\!\left(\frac{\rho_s}{2} \right) 
			+ I_{k+1}\!\!\left(\frac{\rho_s}{2} \right) \right] } \nonumber \\
&&  \qquad  \times \Re \left\{ \Psi_{\Phi}
                \left[ \frac{(2 k + 1) <\!\!\Phi_{\mathrm{NL}}\!\!> }{ \rho_s + \frac{1}{2}} \right] e^{-j(2k+1) \theta_c }  \right\}. \qquad 
\label{BerApproxPSK}
\end{eqnarray}

\noindent In \cite{ho0309a}, the center phase of $\theta_c$ of \eqn{BerApproxPSK} is assumed to be the mean nonlinear phase shift $<\!\!\Phi_{\mathrm{NL}}\!\!>$.

\begin{figure}
\centerline{\includegraphics[width = 0.33 \textwidth]{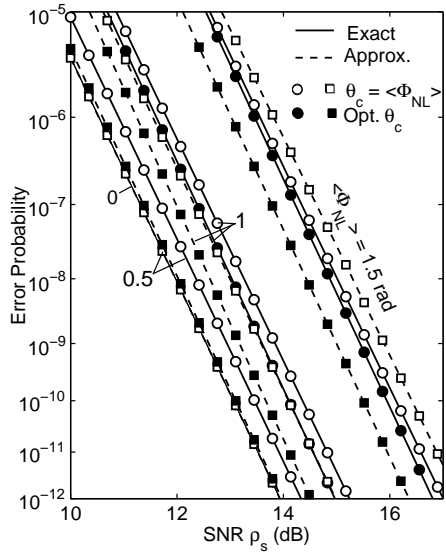}}
\caption{The error probability of PSK signal $p_{e, \mathrm{PSK}}$ as a function of SNR $\rho_s$.}
\label{figBerPSK}
\end{figure}

\fig{figBerPSK} shows the exact \eqn{BerPSK} and approximated \eqn{BerApproxPSK} error probabilities as a function of SNR $\rho_s$.
\fig{figBerPSK} also plots the error probability without nonlinear phase noise of $p_{e, \mathrm{PSK}} = \frac{1}{2}\mathrm{erfc} \sqrt{\rho_s}$ \cite[Sec. 5.2.7]{proakis4}.
\fig{figBerPSK} plots the error probability for both the center phase equal to the mean nonlinear phase shift $\theta_c = <\!\!\Phi_{\mathrm{NL}}\!\!>$ (empty symbol) \cite{mecozzi94} and optimized to minimize the error probability (solid symbol). 
The approximated error probability in \fig{figBerPSK} with $\theta_c = <\!\!\Phi_{\mathrm{NL}}\!\!>$ is the same as that in \cite{ho0309a} but calculated by a simple formula of \eqn{BerApproxPSK}.
From \fig{figBerPSK}, with optimized center phase, the approximated error probability \eqn{BerApproxPSK} always underestimates the error probability.

\begin{figure}
\centerline{\includegraphics[width = 0.4 \textwidth]{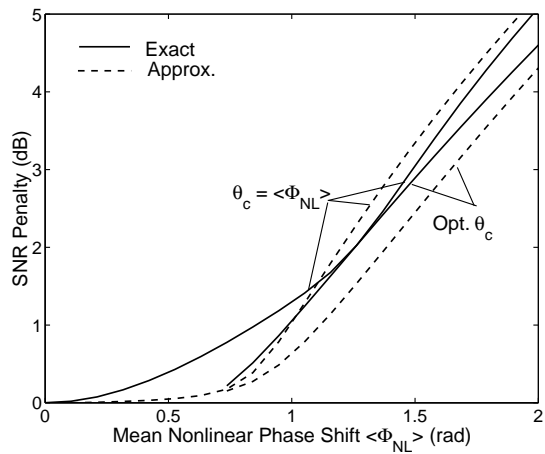}}
\caption{The SNR penalty of PSK signal as a function of mean nonlinear phase shift $<\!\!\Phi_{\mathrm{NL}}\!\!>$.}
\label{figPenPSK}
\end{figure}

\begin{figure}
\centerline{\includegraphics[width = 0.4 \textwidth]{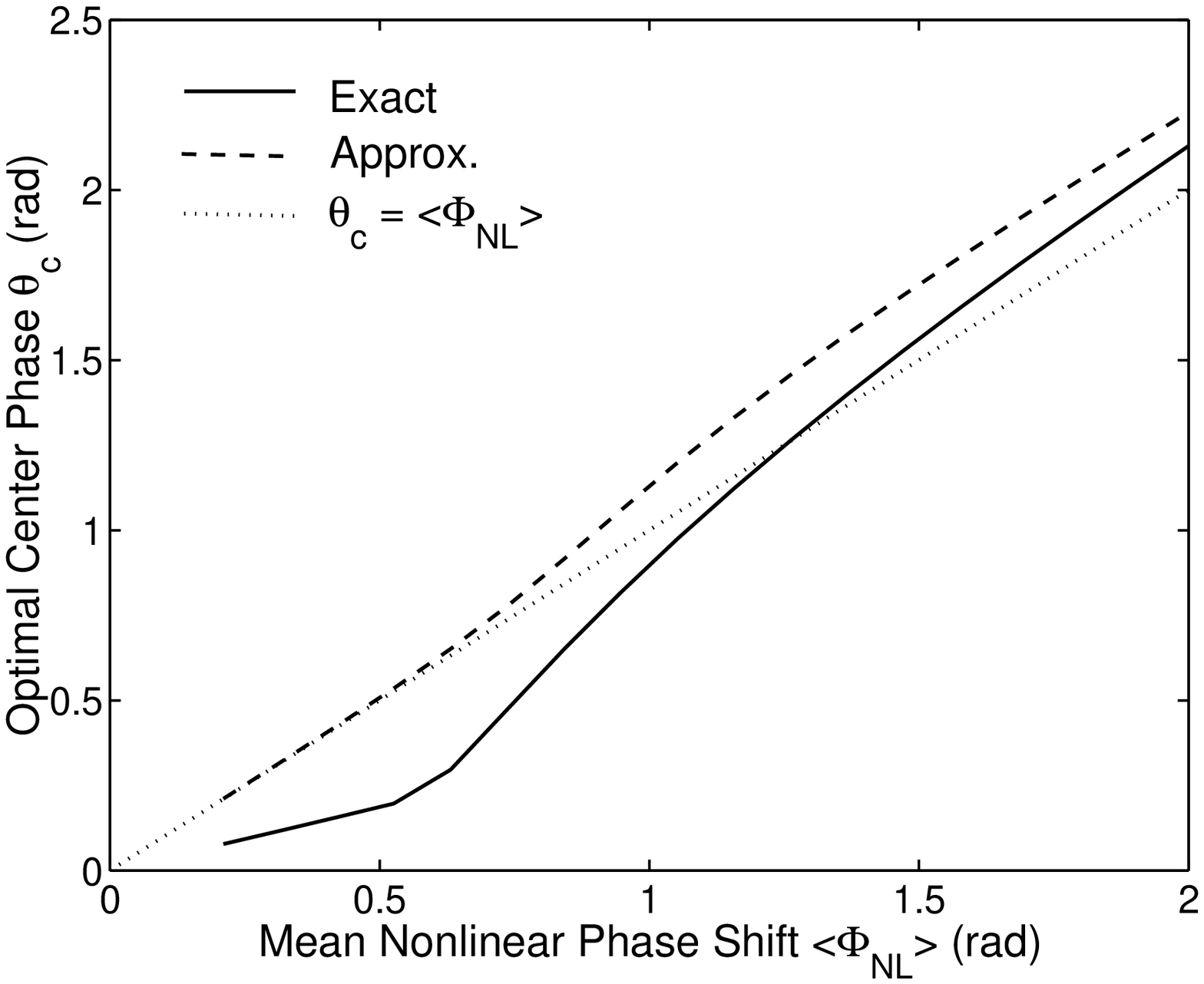}}
\caption{The optimal center phase corresponding to the operating point of \fig{figPenPSK} as a function of mean nonlinear phase shift $<\!\!\Phi_{\mathrm{NL}}\!\!>$.}
\label{figOptTheta}
\end{figure}

\fig{figPenPSK} shows the SNR penalty of PSK signal for an error probability of $10^{-9}$ calculated by the exact \eqn{BerPSK} and approximated \eqn{BerApproxPSK} error probability formulae.
\fig{figPenPSK} is plotted for both cases of the center phase equal to the mean nonlinear phase shift $\theta_c = <\!\!\Phi_{\mathrm{NL}}\!\!>$ or optimized to minimize the error probability. 
The corresponding optimal center phase is shown in \fig{figOptTheta}.
When the center phase is equal to the mean nonlinear phase shift, the results using the exact error probability \eqn{BerPSK} should be the similar to that of \cite{mecozzi94}.
When the center phase is equal to the mean nonlinear phase shift $\theta_c = <\!\!\Phi_{\mathrm{NL}}\!\!>$, the SNR penalty given by the approximated error probability \eqn{BerApproxPSK} is the same as that in \cite{ho0309a} but calculated by a simple formula.

The discrepancy between the exact and approximated error probability is smaller for small and large nonlinear phase shift $<\!\!\Phi_{\mathrm{NL}}\!\!>$.
With the optimal center phase, the largest discrepancy between the exact and approximated SNR penalty is about 0.49 dB at a mean nonlinear phase shift of $<\!\!\Phi_{\mathrm{NL}}\!\!>$ around 1.25 rad. 
When the center phase is equal to the mean nonlinear phase shift  $\theta_c = <\!\!\Phi_{\mathrm{NL}}\!\!>$, the largest discrepancy between the exact and approximated SNR penalty is about 0.6 dB at a mean nonlinear phase shift of $<\!\!\Phi_{\mathrm{NL}}\!\!>$ around 0.75 rad. 
For PSK signal, the approximated error probability \eqn{BerApproxPSK} may not accurate enough for practical applications.

Using the exact error probability \eqn{BerPSK} with optimal center phase, the mean nonlinear phase shift $<\!\!\Phi_{\mathrm{NL}}\!\!>$ must be less than 1 rad for a SNR penalty less than 1 dB. 
The optimal operating level is that the increase of mean nonlinear phase shift, proportional to the increase of launched power and SNR, does not decrease the system performance.
In \fig{figPenPSK}, the optimal operation point can be found by 

\begin{equation}
\frac{d \rho_s}{d <\!\!\Phi_{\mathrm{NL}}\!\!>} \rightarrow 1 
\end{equation}

\noindent when both the required SNR $\rho_s$ and mean nonlinear phase shift $<\!\!\Phi_{\mathrm{NL}}\!\!>$ are expressed in decibel unit.
The optimal operating level is for the mean nonlinear phase noise $<\!\!\Phi_{\mathrm{NL}}\!\!>$ of about 1.25 rad, close to the estimation of \cite{mecozzi94} when the center phase is assumed to be $<\!\!\Phi_{\mathrm{NL}}\!\!>$.

From the optimal center phase of \fig{figOptTheta} with the exact error probability \eqn{BerPSK}, the optimal center phase is less than the mean nonlinear phase shift $<\!\!\Phi_{\mathrm{NL}}\!\!>$ when the mean nonlinear phase shift is less than about 1.25 rad.
At small mean nonlinear phase shift, from \fig{figpdfPhiR}, the p.d.f. of the received phase spreads further to positive phase such that the optimal center phase is smaller that the mean nonlinear phase shift $<\!\!\Phi_{\mathrm{NL}}\!\!>$.
At large mean nonlinear phase shift $<\!\!\Phi_{\mathrm{NL}}\!\!>$, the received phase is dominated by the nonlinear phase noise.
Because the p.d.f. of nonlinear phase noise spreads further to the negative phase as from \fig{pdfsnr}, the optimal center phase is larger than the mean nonlinear phase shift for large mean nonlinear phase shift.
For the same reason, when the nonlinear phase noise is assumed to be independent of the phase of amplifier noise, the optimal center phase is always larger than the mean nonlinear phase shift.
From \fig{figOptTheta}, the approximated error probability \eqn{BerApproxPSK} is not useful to find the optimal center phase.

Comparing the exact \eqn{BerPSK} and approximated \eqn{BerApproxPSK} error probability, the approximated error probability \eqn{BerApproxPSK} is evaluated when the parameter of $\lambda_k$ is approximated by the SNR $\rho_s$.
The parameters of $\lambda_k$ are complex numbers. 
Because $|\lambda_k|$ are always less than $\rho_s$, with optimized center phase and from \figs{figBerPSK} and \ref{figPenPSK}, the approximated error probability of \eqn{BerApproxPSK} always gives an error probability smaller than the exact error probability \eqn{BerPSK}.

\subsection{Error Probability of DPSK Signals}

\begin{figure}
\centerline{\includegraphics[width = 0.45 \textwidth]{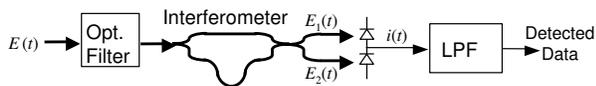}}
\caption{The direct-detection receiver of DPSK signal using an interferometer.}
\label{figRX}
\end{figure}

\fig{figRX} shows the direct-detection receiver for DPSK signal. 
The DPSK receiver uses a Mach-Zehnder interferometer in which the signal is splitted into two paths and combined with a path difference of a symbol time of $T$.
In practice, the path difference $\tau \approx T$ must be chosen such that $\exp(j \omega_0 \tau) = 1$, where $\omega_0$ is the angular frequency of the signal \cite{swanson94, blachman81, rohde00, poggiolini02, kim03freq, bosco03, winzer03}.
Ideally, the optical filter before the interferometer is assumed to be a match filter to the transmitted signal.
Two balanced photodetectors are used to receive the photocurrent. 
There is a low-pass filter to filter out the receiver noise.
We assume that the low-pass filter has a wide bandwidth and does not distort the received signal.

Optical amplified direct-detection DPSK receiver had been studied by \cite{humblet91, tonguz91, pires92, chinn96}.
The analysis here just takes into account the amplifier noise from the same polarization as the signal \cite{chinn96} and can also be applied to heterodyne receiver \cite{tonguz91}.

The interferometer of \fig{figRX} finds the differential phase of

\begin{eqnarray}
\lefteqn {\Delta \Phi_r = \Phi_r(t) - \Phi_r(t-T) } \nonumber \\
& &  = \Theta_n(t) - \Phi_{\mathrm{NL}}(t)
		 - \Theta_n(t-T) + \Phi_{\mathrm{NL}}(t-T) ,	 
\label{DeltaPhiR}
\end{eqnarray}

\noindent where $\Phi_r(\cdot)$, $\Theta_n(\cdot)$, and $\Phi_{\mathrm{NL}}(\cdot)$ are the received phase, the phase of amplifier noise, and the nonlinear phase noise as a function of time, and $T$ is the symbol interval.
The phases at $t$ and $t-T$ are independent of each other but are identically distributed random variables similar to that of \eqn{PhiR}. 
The differential phase of \eqn{DeltaPhiR} assumes that the transmitted phases at $t$ and $t-T$ are the same.

When two random variables are added (or subtracted) together, the sum has a characteristic function that is the product of the corresponding individual characterstic functions.
The p.d.f. of the sum of the two random variables has Fourier series coefficients that are the product of the corresponding Fourier series coefficients.
From \eqn{pdfPhiR}, the p.d.f. of the differential phase \eqn{DeltaPhiR} is

\begin{equation}
p_{\Delta \Phi_r}(\theta) = \frac{1}{2 \pi} + \frac{1}{\pi} 
\sum_{m =1}^{+\infty} \left| \Psi_{\Phi_r}(m) \right|^2 \cos( m \theta).
\label{pdfDeltaPhiR}
\end{equation}

As the difference of two i.i.d. random variables, with the same transmitted phase in two consecutive symbols, the p.d.f. of the differential phase $\Delta \Phi_r$ is symmetrical with respect to the zero phase. 

\begin{figure}
\centerline{\includegraphics[width = 0.4 \textwidth]{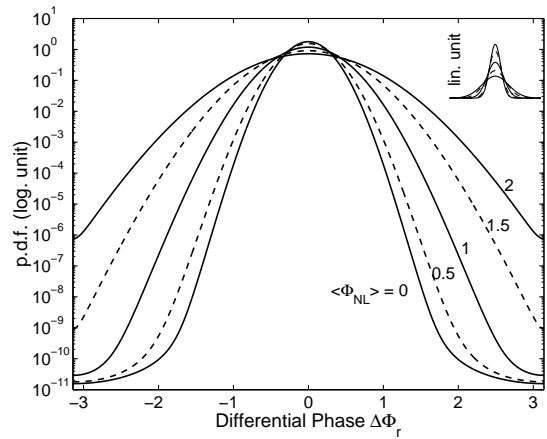}}
\caption{The p.d.f. of the differential received phase $p_{\Delta \Phi_r}(\theta)$ in logarithmic scale. 
The inset is the same p.d.f. in linear scale.}
\label{figpdfDeltaPhiR}
\end{figure}

\fig{figpdfDeltaPhiR} shows the p.d.f. of the differential received phase \eqn{pdfDeltaPhiR} with mean nonlinear phase shift of $<\!\!\Phi_{\mathrm{NL}}\!\!> = 0, 0.5, 1.0, 1.5$, and $2.0$ rad.
The p.d.f. is plotted in logarithmic scale to show the difference in the tail.
The same p.d.f. is plotted in linear scale in the inset.
\fig{figpdfDeltaPhiR} is plotted for the case that the SNR is equal to $\rho_s = 20$ (13 dB), corresponding to an error probability of $10^{-9}$ if amplifier noise is the sole impairment \cite{tonguz91}.
From \fig{figpdfDeltaPhiR}, when the p.d.f. of differential phase is broadened by the nonlinear phase noise, the broadening is symmetrical with respect to the zero phase.

Interferometer based receiver \cite{humblet91, pires92, chinn96} gives an output proportional to $\cos(\Delta \Phi_r)$.
The detector makes a decision on whether $\cos(\Delta \Phi_r)$ is positive or negative that is equivalent to whether the differential phase $\Delta \Phi_r$ is within or without the angle of $\pm \frac{1}{2} \pi$.
Similar to that for PSK signal \eqn{BerPSK}, the error probability for DPSK signal is

\begin{equation}
p_{e, \mathrm{DPSK}} = \frac{1}{2} - \frac{2}{\pi}
\sum _{k =0}^{+\infty}
   \frac{(-1)^k}{2 k + 1} \left| \Psi_{\Phi_r}(2 k + 1) \right|^2.
\label{BerDPSK}
\end{equation}

\noindent where the coefficients of $ \Psi_{\Phi_r}(2 k + 1)$ are given by \eqn{cfPhiRk}.

Similar to the approximation for PSK signal \eqn{BerApproxPSK}, if the nonlinear phase noise is assumed to be independent to the phase of amplifier noise, the error probability of \eqn{BerDPSK} can be approximated as

\begin{eqnarray}
\lefteqn{ p_{e, \mathrm{DPSK}} \approx \frac{1}{2} - \frac{\rho_s e^{-\rho_s}}{2} 
	  \sum_{k = 0}^{\infty} \frac{(-1)^k} {2 k + 1}
\left[ I_{k}\!\!\left(\frac{\rho_s}{2} \right) 
			+ I_{k+1}\!\!\left(\frac{\rho_s}{2} \right) \right]^2 } \nonumber \\
&& \qquad \qquad \qquad \quad \times \left| \Psi_{\Phi} 
		\left[ \frac{(2 k + 1) <\!\!\Phi_{\mathrm{NL}}\!\!> }{ \rho_s + \frac{1}{2}} \right]  \right|^2. \qquad
\label{BerApproxDPSK}
\end{eqnarray}

Comparing the exact \eqn{BerDPSK} and approximated \eqn{BerApproxDPSK} error probability, the approximated error probability \eqn{BerApproxDPSK} is evaluated when the parameter of $\lambda_k$ is approximated by the SNR $\rho_s$.
Because $|\lambda_k|$ is always less than $\rho_s$, the approximated error probability of \eqn{BerApproxDPSK} always gives an error probability smaller than the exact error probability \eqn{BerDPSK}.

\begin{figure}
\centerline{\includegraphics[width = 0.33 \textwidth]{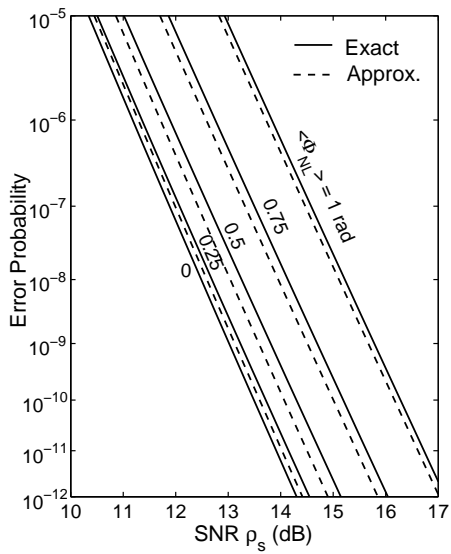}}
\caption{The error probability of DPSK signal as a function of SNR $\rho_s$.}
\label{figBerDPSK}
\end{figure}

\fig{figBerDPSK} shows the exact \eqn{BerDPSK} and approximated \eqn{BerApproxDPSK} error probabilities as a function of SNR $\rho_s$.
\fig{figBerDPSK} also plots the error probability without nonlinear phase noise of $p_{e, \mathrm{DPSK}} = \frac{1}{2} \exp(-\rho_s)$ \cite[Sec. 5.2.8]{proakis4}.
The approximated error probability in \fig{figBerDPSK} is the same as that in \cite{ho0309a} but calculated by a simple formula of \eqn{BerApproxDPSK}.
From \fig{figBerDPSK}, the approximated error probability \eqn{BerApproxDPSK} always underestimates the error probability.

\begin{figure}
\centerline{\includegraphics[width = 0.4 \textwidth]{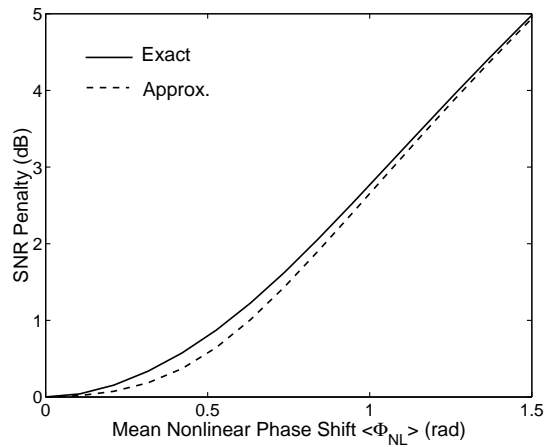}}
\caption{The SNR penalty of DPSK signal as a function of mean nonlinear phase shift $<\!\!\Phi_{\mathrm{NL}}\!\!>$.}
\label{figPenDPSK}
\end{figure}

\fig{figPenDPSK} shows the SNR penalty of DPSK signal for an error probability of $10^{-9}$ calculated by the exact \eqn{BerDPSK} and approximated \eqn{BerApproxDPSK} error probability formulae.
The SNR penalty given by the approximated error probability is the same as that in \cite{ho0309a} but calculated by a simple formula \eqn{BerApproxDPSK}.
The discrepancy between the exact and approximated error probability is very small for small and large nonlinear phase shift $<\!\!\Phi_{\mathrm{NL}}\!\!>$.
The largest discrepancy between the exact and approximated SNR penalty is about 0.27 dB at a mean nonlinear phase shift $<\!\!\Phi_{\mathrm{NL}}\!\!>$ of $0.53$ rad. 

For a power penalty less than 1 dB, the mean nonlinear phase shift  $<\!\!\Phi_{\mathrm{NL}}\!\!>$ must be less than 0.57 rad. 
The optimal level of the mean nonlinear phase shift  $<\!\!\Phi_{\mathrm{NL}}\!\!>$ is about 1 rad such that the increase of power penalty is always less than the increase of mean nonlinear phase shift,  similar to the estimation of \cite{gordon90} as the limitation of the mean nonlinear phase shift.

The error probabilities of \figs{figBerPSK} and \ref{figBerDPSK} are calculated using Matlab.
The series summation of \eqn{BerPSK} and \eqn{BerDPSK} can be calculated to an error probability of $10^{-13}$ to $10^{-14}$ with an accuracy of three to four significant digits.
Symbolic mathematical software can provide better accuracy by using variable precision arithmetic in the calculation of low error probability.

\section{Linear Compensation of Nonlinear Phase Noise}
\label{sec:lincomp}

As shown in the helix shape scattergram of \figs{figyinyang}, nonlinear phase noise is correlated with the received intensity. 
The received intensity can be used to compensate for the nonlinear phase noise.
Ideally, as shown in coming sections, the optimal compensator should minimize the error probability of the system after compensation. 
If the joint p.d.f.~of the received phase and the received intensity of $p_{\Phi_r, Y}(\theta, y)$ is available, the optimal compensator or detector is given by the maximum {\em a posteriori} probability (MAP) criterion \cite[Sec. 5.2]{mcdonough2} to minimize the error probability.
For simplicity, linear compensator is discussed here first.
The next section is about nonlinear compensator.

In this section, the linear compensator is first optimized in term of the variance, or minimum mean-square error (MMSE) criterion, of the residual nonlinear phase noise.
Afterward, the exact error probability with linear compensator is derived analytically.
With a simple expression to calculate the error probability, numerical optimization is used to find the linear MAP compensator to minimize the error probability.
Not for PSK signals, linear MMSE compensator performs close to linear MAP compensator for DPSK signals.

\subsection{MMSE Linear Compensation}

The simplest method to compensate the nonlinear phase noise is to add a scaled received intensity into the received phase \cite{liu02, xu02, xu02a, ho0403a}. 
The optimal linear MMSE compensator minimizes the variance of the normalized residual phase noise of $\Phi_\alpha = \Phi - \alpha R^2 = \Phi - \alpha Y$.
Using the joint characteristic function of \eqn{cfPhiY}, the characteristic function for the normalized residual nonlinear phase noise  is 

\begin{equation}
\Psi_{\Phi_\alpha}(\nu) =  \Psi_{\Phi, Y}(\nu, - \alpha \nu).
\end{equation}

\noindent The mean of the normalized residual nonlinear phase noise is 

\begin{eqnarray}
<\!\!\Phi_\alpha\!\!> & = & - j \left. \frac{ d}{d \nu}  \Psi_{\Phi_\alpha}(\nu) \right|_{\nu = 0} \nonumber \\
	&=&  \rho_s+ \frac{1}{2} - \alpha \left( \rho_s + 1\right).
\end{eqnarray}

\noindent  The variance of the normalized residual nonlinear phase noise is

\begin{eqnarray}
\lefteqn{ \sigma_{\Phi_\alpha}^2  =  - \left. \frac{ d^2}{d \nu^2}  \Psi_{\Phi_\alpha}(\nu) \right|_{\nu = 0} - <\!\!\Phi_\alpha\!\!>^2} \nonumber \\
	&&=  \frac{2}{3}\rho_s + \frac{1}{6}  - 
	2 \left(\rho_s + \frac{1}{3}\right) \alpha  +\left(2\rho_s + 1\right)  \alpha^2 .
\end{eqnarray}

\noindent Solving $d \sigma_{\Phi_\alpha}^2/d \alpha = 0$, the optimal scale factor for linear compensator is

\begin{equation}
\alpha_{\mathrm{min}} = \frac{1}{2} \frac{ \rho_s + \frac{1}{3}}{ \rho_s + \frac{1}{2}}.
\label{alphaopt}
\end{equation}

\noindent In high SNR, $\alpha_{\mathrm{min}} \rightarrow \frac{1}{2}$.
Other than the normalization, the optimal scale factor of \eqn{alphaopt} is the same as that in \cite{ho0403a}. 
The approximation of $\alpha_{\mathrm{min}} \rightarrow \frac{1}{2}$ was estimated by \cite{xu02} though simulation.

With the optimal scale factor of \eqn{alphaopt}, the mean and variance of the normalized residual nonlinear phase noise are

\begin{eqnarray}
<\!\!\Phi_{\alpha_{\mathrm{min}}}\!\!> &=& \frac{1}{2} \frac{\rho_s^2 + \frac{2}{3}\rho_s + \frac{1}{6}}{\rho_s + \frac{1}{2}}, 
\label{meanmin} \\
 \sigma_{\Phi_{\alpha_{\mathrm{min}}}}^2 & = & \frac{1}{6} \frac{\rho_s^2+ \rho_s + \frac{1}{6}}{\rho_s + \frac{1}{2}}.
\label{sigmares}
\end{eqnarray}

\noindent The mean of the residual nonlinear phase noise is about half the mean of the nonlinear phase noise of $<\!\!\Phi\!\!>$ \eqn{meanPhi}.
The variance of the residual nonlinear phase noise is about a quarter of that of the variance of the nonlinear phase noise of $\sigma_{\Phi}^2$ \eqn{sigmaPhi}.

After the linear compensation, the characteristic function of the residual normalized nonlinear phase noise is

\begin{equation}
\Psi_{\Phi_{\alpha_{\mathrm{min}}}}(\nu) =  \Psi_{\Phi, Y}\!\left(\nu, - \frac{\nu}{2} \frac{\rho_s + \frac{1}{3}}{\rho_s+ \frac{1}{2}}\right).
\label{cfPhiAlpha}
\end{equation}

\noindent The p.d.f.~of the residual normalized nonlinear phase noise is the inverse Fourier transform of $\Psi_{\Phi_{\alpha_{\mathrm{min}}}}\!(\nu)$.

\begin{figure}
\centerline{\includegraphics[width = 0.4 \textwidth]{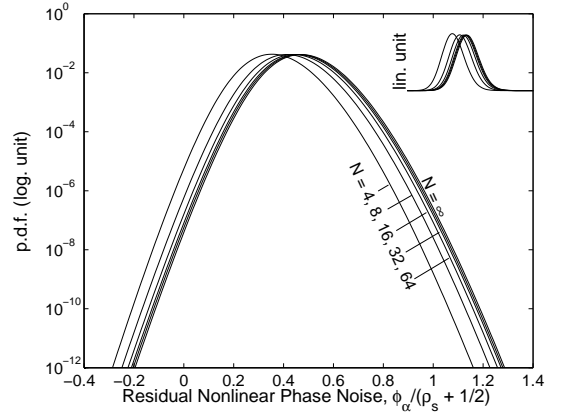}}
\caption{The asymptotic p.d.f.~of the residual nonlinear phase noise $\Phi_\alpha$ as compared with the p.d.f.~of $N=4, 8, 16, 32,$ and $64$ fiber spans.
The p.d.f.~in linear scale is shown in the inset.}
\label{pdfNcomp}
\end{figure}

The p.d.f.~of the residual nonlinear phase noise for finite number of fiber spans was derived by modeling the residual nonlinear phase noise $\Phi_\alpha$ as the summation of $N$ independently distributed random variables \cite{ho0309c}.  
\fig{pdfNcomp} shows a comparison of the p.d.f.~for  $N = 4, 8, 16, 32$, and $64$ of fiber spans \cite{ho0309c} with the distributed case of \eqn{cfPhiAlpha}. 
The residual nonlinear phase noise is scaled by the mean normalized phase shift of $<\!\!\Phi\!\!> = \rho_s + \frac{1}{2}$ \eqn{meanPhi}.
Using a SNR of $\rho_s = 18$, \fig{pdfNcomp} is plotted in logarithmic scale to show the difference in the tail.
\fig{pdfNcomp} also provides an inset in linear scale of the same p.d.f.~to show the difference around the mean.
Like that of \fig{pdfN}, the asymptotic p.d.f.~for residual nonlinear phase noise of \fig{pdfNcomp} is also very accurate for $N \geq 32$ fiber spans.  
Unlike that of \fig{pdfN}, the asymptotic p.d.f.~for residual nonlinear phase noise of \fig{pdfNcomp} have slightly larger spread then that of the finite cases. 
The mean of the residual nonlinear phase noise is about $0.5 (\rho_s + \frac{1}{2})$, the same as that of \eqn{meanmin}.
Comparing \fig{pdfN} and \fig{pdfNcomp}, with linear compensation, both the mean and STD of the residual nonlinear phase noise is about half of that of the nonlinear phase noise before compensation.

\subsection{Distribution of the Linearly Compensated Received Phase}

With nonlinear phase noise, received by an optical phase-locked loop \cite{kahn90, norimatsu92}, the overall received phase is that of \eqn{PhiR}.
With linear compensation, the compensated received phase is

\begin{equation}
\Phi_{\mathrm{cm}} = \Theta_n - \frac{ <\!\! \Phi_{\mathrm{NL}}\!\!>} {\rho_s + \frac{1}{2}} \left(\Phi -\alpha Y \right) ,
\label{PhiCM}
\end{equation}

The compensated received phase \eqn{PhiCM} is confined to the range of $[-\pi, +\pi)$. 
The p.d.f. of the compensated received phase is a periodic function with a period of $2 \pi$.
If the characteristic function of the compensated received phase is $\Psi_{\Phi_{\mathrm{cm}}}(\nu)$, the p.d.f. of the compensated received phase has a Fourier series expansion of

\begin{equation}
p_{\Phi_{\mathrm{cm}}}(\theta) =\frac{1}{2 \pi} \sum_{m = -\infty}^{+\infty} \Psi_{\Phi_{\mathrm{cm}}}(m) \exp( j m \theta).
\label{pdfPhiCM1}
\end{equation}

\noindent or

\begin{equation}
p_{\Phi_{\mathrm{cm}}}(\theta) = \frac{1}{2 \pi} + \frac{1}{\pi} 
\sum_{m =1}^{+\infty}\Re \left\{ \Psi_{\Phi_{\mathrm{cm}}}(m) \exp( j m \theta) \right\},
\label{pdfPhiCM}
\end{equation}

\noindent where $\Re\{\cdot\}$ denotes the real part of a complex number.

Using the joint characteristic function of $\Psi_{\Phi, Y, \Theta_n}(\nu, \omega, m)$ \eqn{cfPhiThetam}, similar to that of \eqn{cfPhiAlpha}, from the compensated received phase of \eqn{PhiCM}, the Fourier series coefficients are

\begin{equation}
 \Psi_{\Phi_{\mathrm{cm}}}(m) = \Psi_{\Phi, Y, \Theta_n} \left( - 
		 \frac{m <\!\! \Phi_{\mathrm{NL}}\!\!>} {\rho_s + \frac{1}{2}}, 
                \alpha \frac{m  <\!\! \Phi_{\mathrm{NL}}\!\!>} {\rho_s + \frac{1}{2}},
		m \right).
\label{cfPhiCM}
\end{equation}

\begin{figure}
\centerline{\includegraphics[width = 0.4 \textwidth]{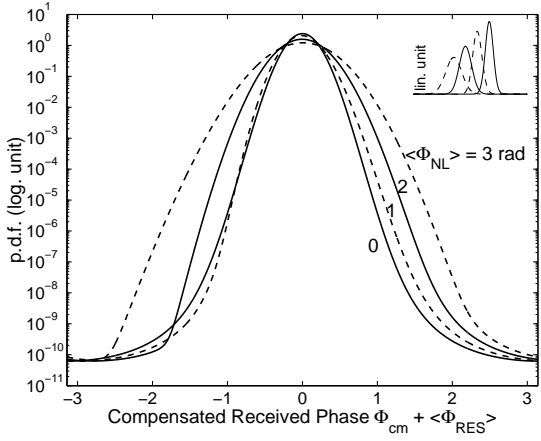}}
\caption{The p.d.f. of the compensated received phase $p_{\Phi_{\mathrm{cm}}}(\theta +<\!\!\Phi_{\mathrm{RES}}\!\!>)$ in logarithmic scale. 
The inset is the p.d.f. of $p_{\Phi_{\mathrm{cm}}}(\theta)$ in linear scale.}
\label{figpdfPhiCM}
\end{figure}

Using the scale factor $\alpha_\mathrm{min}$ \eqn{alphaopt}, \fig{figpdfPhiCM} shows the p.d.f. of the compensated received phase \eqn{pdfPhiCM} with mean nonlinear phase shift of $<\!\!\Phi_{\mathrm{NL}}\!\!> = 0, 1, 2$, and $3$ rad.
Shifted by the mean residual nonlinear phase shift of

\begin{equation}
<\!\!\Phi_{\mathrm{RES}}\!\!> = \frac{<\!\!\Phi_{\mathrm{NL}}\!\!>}{\rho_s + \frac{1}{2}} <\!\! \Phi_{\alpha_\mathrm{min}}\!\!>,
\label{meanRES}
\end{equation}

\noindent the p.d.f. is plotted in logarithmic scale to show the difference in the tail.
Not shifted by $<\!\!\Phi_{\mathrm{RES}}\!\!>$, the same p.d.f. is plotted in linear scale in the inset.
\fig{figpdfPhiCM} is plotted for the case that the SNR is equal to $\rho_s = 18$ (12.6 dB), the same as that of \fig{figpdfPhiR}.

From \fig{figpdfPhiCM}, when the p.d.f. is broadened by the nonlinear phase noise, similar to the case without compensation of \fig{figpdfPhiR}, the broadening is not symmetrical with respect to the mean residual nonlinear phase shift $<\!\!\Phi_{\mathrm{RES}}\!\!>$ \eqn{meanRES}. 
With small mean nonlinear phase shift of  $<\!\!\Phi_{\mathrm{NL}}\!\!> = 1$ rad, the received phase spreads further in the positive phase than the negative phase.
With large mean nonlinear phase shift of  $<\!\!\Phi_{\mathrm{NL}}\!\!> = 3$ rad, the received phase spreads further in the negative phase than the positive phase.
The difference in the spreading for small and large mean nonlinear phase shift is due to the dependence between the residual nonlinear phase noise and the phase of amplifier noise.

\subsection{Error Probability of PSK Signals}

Assume that the decision angles are $\pm \frac{1}{2} \pi - \theta_c$ with the center phase of $\theta_c$, the error probability is

\begin{equation}
p_{e, \mathrm{PSK}} = 1 - \int_{-\frac{1}{2}\pi -\theta_c}^{ \frac{1}{2}\pi-\theta_c} p_{\Phi_{\mathrm{cm}}} (\theta) \ud \theta,
\end{equation} 

\noindent or, similar to the error probability of \eqn{BerPSK}, 

\begin{equation}
p_{e, \mathrm{PSK}} = \frac{1}{2} - \frac{2}{\pi}
\sum _{k =0}^{+\infty}
   \frac{(-1)^k}{2 k + 1}\Re \left\{ \Psi^*_{\Phi_{\mathrm{cm}}}(2 k + 1) e^{ -j (2 k +1) \theta_c} \right\}.
\label{BerPSKLC}
\end{equation}

\noindent From both \eqn{cfPhiCM} and \eqn{cfPhiYThetam}, the coefficients for the error probability \eqn{BerPSKLC} are

\begin{eqnarray}
\lefteqn{\Psi^*_{\Phi_{\mathrm{cm}}}(2k+1)  =   \frac{\sqrt{\pi} \lambda_{k,\omega}^{\frac{3}{2}}}{2 \lambda_k} e^{-\lambda_k + \frac{\lambda_{k, \omega}}{2}}
\Psi_{\Phi} 
	\left[ \frac{(2 k + 1) <\!\!\Phi_{\mathrm{NL}}\!\!> }{ \rho_s + \frac{1}{2}} \right] } \nonumber \\
&& \qquad  \times \left[I_k \left( \frac{\lambda_{k, \omega}}{2} \right) 
	+ I_{k+1} \left( \frac{\lambda_{k, \omega}}{2} \right) \right], \quad   k \geq 0, \qquad \nonumber \\
\label{cfPhiCMk}
\end{eqnarray}

\noindent where, from \eqn{gammanu} and \eqn{gammano},

\begin{equation}
\lambda_k  = \frac{2 \left[ \frac{j(2 k + 1) <\!\Phi_{\mathrm{NL}}\!> }{ \rho_s + \frac{1}{2}} \right]^{\frac{1}{2}}  }
 	{\sin \left\{2 \left[\frac{j(2 k + 1) <\!\Phi_{\mathrm{NL}}\!> }{ \rho_s + \frac{1}{2}} \right]^{\frac{1}{2}} \right\} } \rho_s, 
\label{lambdakcm}
\end{equation}

\begin{equation}
\lambda_{k, \omega}  = \frac{\lambda_k}{1 +  
\alpha \!\!\left[ \frac{j(2 k + 1) <\!\Phi_{\mathrm{NL}}\!> }{ \rho_s + \frac{1}{2}} \right]^{\frac{1}{2}}\!\!
\tan \left\{\! \left[\frac{j(2 k + 1) <\!\Phi_{\mathrm{NL}}\!> }{ \rho_s + \frac{1}{2}} \right]^{\frac{1}{2}} \! \right\}},
\label{lambdakw}
\end{equation}

\noindent  and $\Psi_\Phi(\nu)$ \eqn{cfPhi} is the marginal characteristic function of nonlinear phase noise  that depends solely on SNR.

Based on the MMSE criterion, the center phase in \eqn{BerPSKLC} is $\theta_c = <\!\!\Phi_\mathrm{RES}\!\!>$ \eqn{meanRES} and the scale factor in \eqn{lambdakw} is $\alpha = \alpha_{\mathrm{min}}$ \eqn{alphaopt}. 
Because the series summation of \eqn{BerPSKLC} is a simple expression, numerical optimization can be used to find the linear MAP compensator to minimize the error probability. 

If the residual nonlinear phase noise is assumed to be independent to the phase of amplifier noise, similar to \eqn{BerApproxPSK}, the error probability based on MMSE criterion can be approximated as

\begin{eqnarray}
\lefteqn{ p_{e, \mathrm{PSK}} \approx \frac{1}{2} - e^{-\frac{\rho_s}{2}}  \sqrt{\frac{\rho_s}{\pi}}
	  \sum_{k = 0}^{\infty} \frac{(-1)^k} {2 k + 1}
\left[ I_{k}\!\!\left(\frac{\rho_s}{2} \right) 
			+ I_{k+1}\!\!\left(\frac{\rho_s}{2} \right) \right] } \nonumber \\
&& \quad \times \Re \left\{ \Psi_{\Phi_{\alpha_\mathrm{min}}}
                \left[ \frac{(2 k + 1) <\!\!\Phi_{\mathrm{NL}}\!\!> }{ \rho_s + \frac{1}{2}} \right] e^{-j(2k+1) <\!\!\Phi_\mathrm{RES}\!\!> }  \right\}. \nonumber \\
\label{BerProxPSKLC}
\end{eqnarray}

\noindent where the characteristic function $\Psi_{\Phi_{\alpha_\mathrm{min}}}(\nu)$ is the characteristic function of \eqn{cfPhiAlpha} with $\alpha = \alpha_{\mathrm{min}}$ \eqn{alphaopt}.

\begin{figure}
\centerline{\includegraphics[width = 0.33 \textwidth]{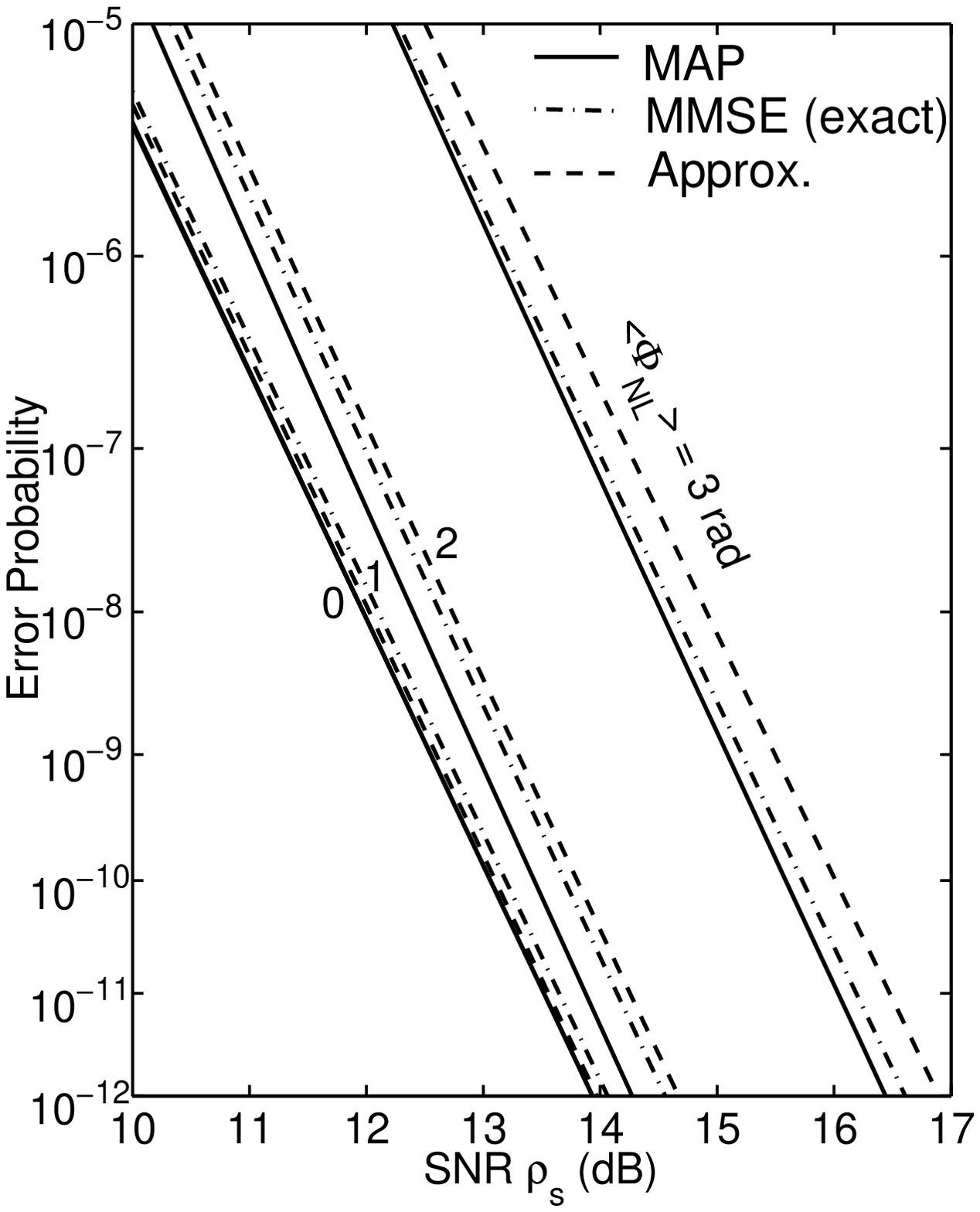}}
\caption{With linear compensation, the error probability of PSK signal $p_{e, \mathrm{PSK}}$ as a function of SNR $\rho_s$.}
\label{figBerPSKLC}
\end{figure}

\fig{figBerPSKLC} shows the exact \eqn{BerPSKLC} and approximated \eqn{BerProxPSKLC} error probabilities as a function of SNR $\rho_s$ when nonlinear phase noise is compensated using the linear MMSE compensator.
The error probability \eqn{BerPSKLC} is also minimized based on the MAP criterion and shown in \fig{figBerPSKLC}.
\fig{figBerPSKLC} also plots the error probability without nonlinear phase noise.

\begin{figure}
\centerline{\includegraphics[width = 0.4 \textwidth]{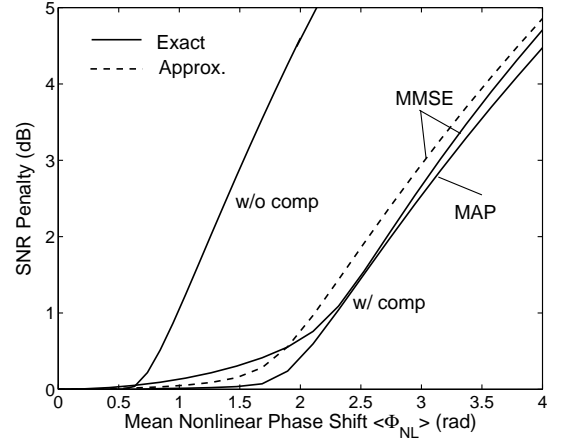}}
\caption{With linear compensation, the SNR penalty of PSK signal as a function of mean nonlinear phase shift $<\!\!\Phi_{\mathrm{NL}}\!\!>$.}
\label{figPenPSKLC}
\end{figure}

\begin{figure}
\centerline{\includegraphics[width = 0.4 \textwidth]{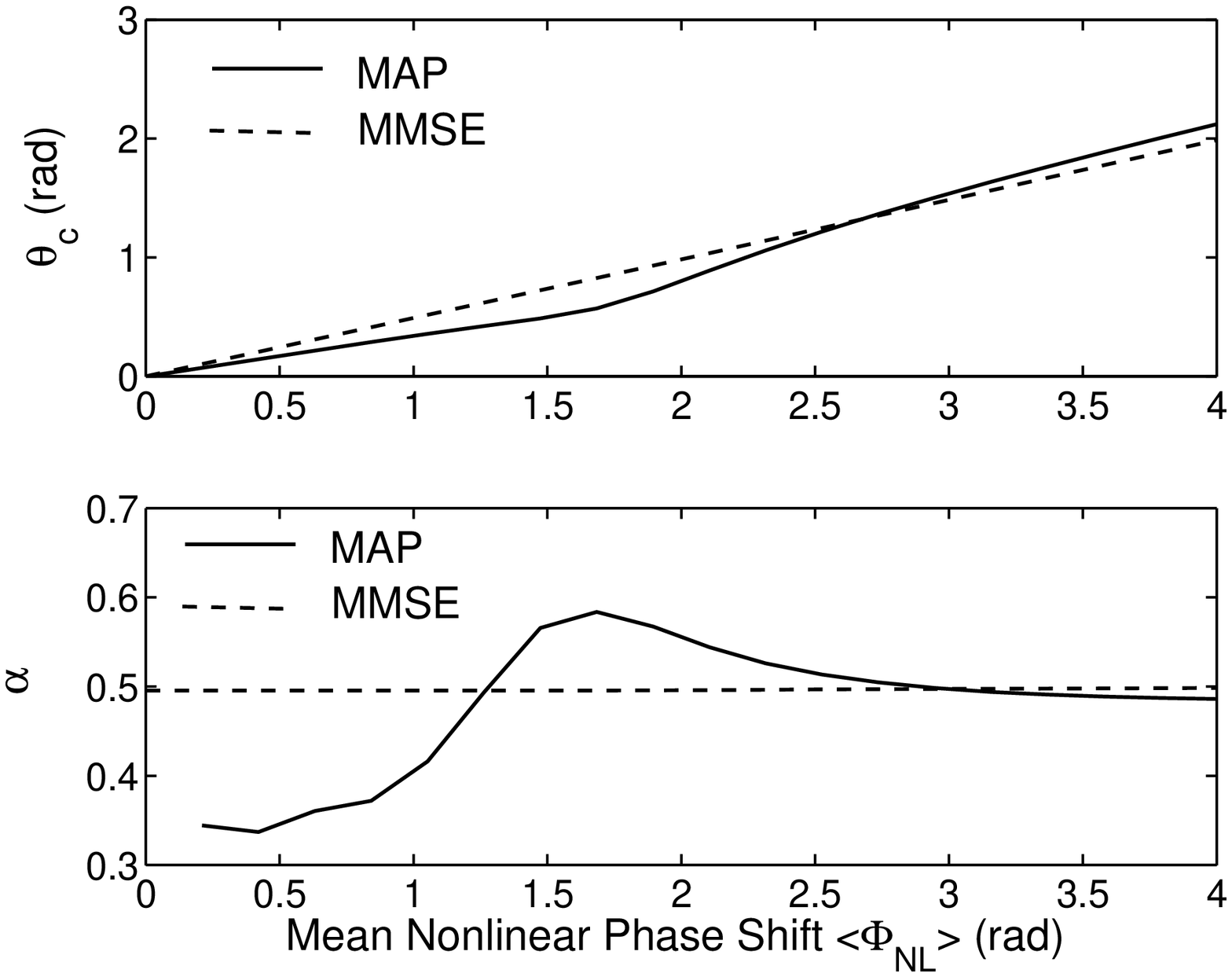}}
\caption{The optimal center phase corresponding to the operating point of \fig{figPenPSKLC} as a function of mean nonlinear phase shift $<\!\!\Phi_{\mathrm{NL}}\!\!>$.}
\label{figOptThetaLC}
\end{figure}

\fig{figPenPSKLC} shows the SNR penalty of PSK signal for an error probability of $10^{-9}$ calculated by the exact \eqn{BerPSKLC} and approximated \eqn{BerProxPSKLC} error probability formulae with the linear MMSE compensator.
The SNR penalty with the linear MAP compensator is also shown in \fig{figPenPSKLC}.
The corresponding optimal center phase and scale factor of \fig{figPenPSKLC} are shown in \fig{figOptThetaLC}.

\fig{figPenPSKLC} also shows the SNR penalty of PSK signal without compensation from \fig{figPenPSK} using the exact error probability with optimal center phase there.
For the same SNR penalty, the mean nonlinear phase shift with compensation is slightly larger than twice of that without compensation from \fig{figPenPSK}. 

The discrepancy between the exact and approximated error probability is smaller for small and large mean nonlinear phase shift $<\!\!\Phi_{\mathrm{NL}}\!\!>$.
With the MMSE criterion, the largest discrepancy between the exact and approximated SNR penalty is about 0.37 dB at a mean nonlinear phase shift $<\!\!\Phi_{\mathrm{NL}}\!\!>$ around 2.53 rad. 
Like the conclusion of Sec. \ref{sec:ber} without compensation, with the MMSE criterion, the approximated error probability \eqn{BerProxPSKLC} is not accurate enough for linearly compensated PSK signals.

\fig{figPenPSKLC} also shows that the linear MMSE compensator using the optimal scale factor of \eqn{alphaopt} does not perform well as compared with the linear MAP compensator.
The largest discrepancy is about 0.34 dB at mean nonlinear phase shift of $<\!\!\Phi_{\mathrm{NL}}\!\!> = 1.68$ rad.
The major reason of this large discrepancy is due to the non-symmetrical p.d.f. of \fig{figpdfPhiCM}.

Using the linear MAP compensator, the mean nonlinear phase shift $<\!\!\Phi_{\mathrm{NL}}\!\!>$ must be less than 2.30 rad for a SNR penalty less than 1 dB, slightly more than twice that of \fig{figPenPSK} of 1 rad without compensation.
The optimal operating level is for a mean nonlinear phase noise $<\!\!\Phi_{\mathrm{NL}}\!\!>$ of about 2.15 rad, slightly less than twice that of \fig{figPenPSK} of 1.25 rad without compensation.

From the optimal center phase from \fig{figOptThetaLC} for the linear MAP compensator, the optimal center phase is less than the mean nonlinear phase shift $<\!\!\Phi_{\mathrm{RES}}\!\!>$ when the mean nonlinear phase shift is less than about 2.68 rad.
The changing of the optimal center phase with the mean nonlinear phase shift is consistent with  \figs{figOptTheta} and \ref{figpdfPhiCM}.
Similar to \fig{figOptTheta}, when the residual nonlinear phase noise is assumed to be independent of the phase of amplifier noise using \eqn{BerProxPSKLC}, the optimal center phase is always larger than the mean residual nonlinear phase shift.
The approximated error probability \eqn{BerProxPSKLC} is not useful to find the optimal center phase. 

The optimal scale factor from \fig{figOptThetaLC} is also not equal to the optimal MMSE scale factor \eqn{alphaopt} except when the mean nonlinear phase shift is very large.
From \figs{figPenPSKLC} and \ref{figOptThetaLC}, the MMSE criterion is not close to the performance of the optimal linearly compensated PSK signals based on MAP criterion. 

\subsection{Error Probability of DPSK Signals}

A compensated DPSK signal is demodulated using interferometer of \fig{figRX} to find the compensated differential phase of

\begin{eqnarray}
\lefteqn {\Delta \Phi_{\mathrm{cm}} = \Phi_{\mathrm{cm}}(t) - \Phi_{\mathrm{cm}}(t-T) } \nonumber \\
& &  = \Theta_n(t) - \Phi_{\mathrm{RES}}(t)
		 - \Theta_n(t-T) + \Phi_{\mathrm{RES}}(t-T), \quad	 
\label{DelPhiCM}
\end{eqnarray}

\noindent where $\Phi_{\mathrm{cm}}(\cdot)$, $\Theta_n(\cdot)$, and $\Phi_{\mathrm{RES}}(\cdot)$ are the compensated received phase, the phase of amplifier noise, and the residual nonlinear phase noise as a function of time, and $T$ is the symbol interval.
The phases at $t$ and $t-T$ are independent of each other but are identically distributed random variables similar to that of \eqn{PhiR} and \eqn{PhiCM}. 
The differential phase of \eqn{DelPhiCM} assumes that the transmitted phases at $t$ and $t-T$ are the same.

From the p.d.f. of \eqn{pdfPhiCM}, similar to \eqn{pdfDeltaPhiR}, the p.d.f. of the differential compensated phase \eqn{DelPhiCM} is

\begin{equation}
p_{\Delta \Phi_{\mathrm{cm}}}(\theta) = \frac{1}{2 \pi} + \frac{1}{\pi} 
\sum_{m =1}^{+\infty} \left| \Psi_{\Phi_{\mathrm{cm}}}(m) \right|^2 \cos( m \theta),
\label{pdfDelPhiCM}
\end{equation}

\noindent that is symmetrical with respect to the zero phase. 

\begin{figure}
\centerline{\includegraphics[width = 0.4 \textwidth]{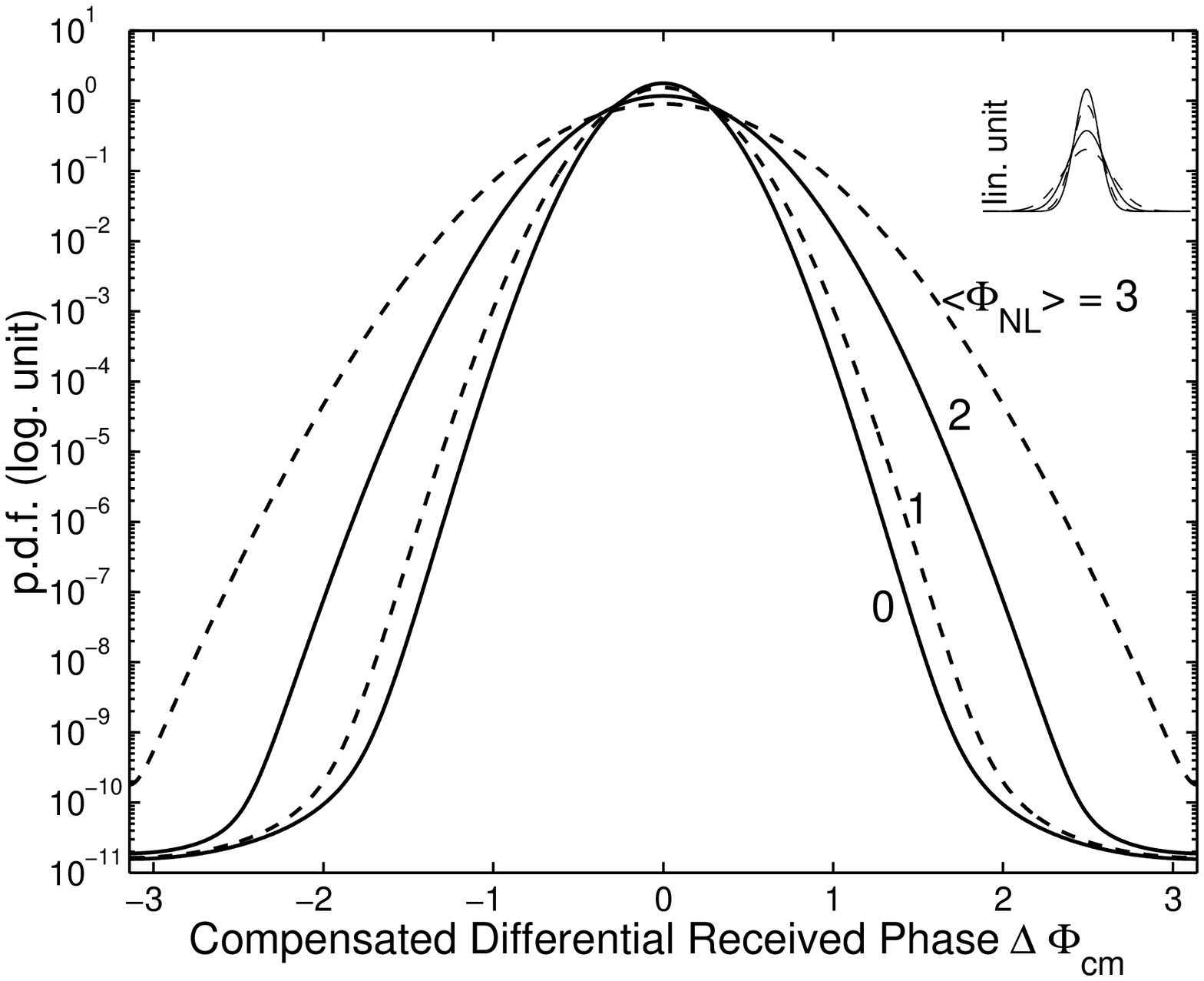}}
\caption{The p.d.f. of the differential compensated received phase $p_{\Delta \Phi_{\mathrm{cm}}}(\theta)$ in logarithmic scale. 
The inset is the same p.d.f. in linear scale.}
\label{figpdfDelPhiCM}
\end{figure}

\fig{figpdfDelPhiCM} shows the p.d.f. of the differential received phase \eqn{pdfDelPhiCM} with mean nonlinear phase shift of $<\!\!\Phi_{\mathrm{NL}}\!\!> = 0,  1, 2$, and $3$ rad using the linear MMSE compensator with $\alpha = \alpha_{\mathrm{min}}$ \eqn{alphaopt}.
The p.d.f. is plotted in logarithmic scale to show the difference in the tail.
The same p.d.f. is plotted in linear scale in the inset.
\fig{figpdfDelPhiCM} is plotted for the case that the SNR is equal to $\rho_s = 20$ (13 dB).
From \fig{figpdfDelPhiCM}, when the p.d.f. of differential phase is broadened by the residual nonlinear phase noise, the broadening is symmetrical with respect to the zero phase.

Similar to \eqn{BerDPSK} and \eqn{BerPSKLC}, the error probability for DPSK signal is

\begin{equation}
p_{e, \mathrm{DPSK}} = \frac{1}{2} - \frac{2}{\pi}
\sum _{k =0}^{+\infty}
   \frac{(-1)^k}{2 k + 1} \left| \Psi_{\Phi_{\mathrm{cm}}}(2 k + 1) \right|^2.
\label{BerDPSKLC}
\end{equation}

\noindent where the coefficients of $ \Psi_{\Phi_{\mathrm{cm}}}(2 k + 1)$ are given by \eqn{cfPhiCMk} with parameters from \eqn{lambdakcm} and \eqn{lambdakw}.

Similar to the approximation for PSK signal \eqn{BerProxPSKLC}, if the nonlinear phase noise is assumed to be independent to the phase of amplifier noise, the error probability of \eqn{BerDPSKLC} can be approximated as

\begin{eqnarray}
\lefteqn{ p_{e, \mathrm{DPSK}} \approx \frac{1}{2} - \frac{\rho_s e^{-\rho_s}}{2} 
	  \sum_{k = 0}^{\infty} \frac{(-1)^k} {2 k + 1}
\left[ I_{k}\!\!\left(\frac{\rho_s}{2} \right) 
			+ I_{k+1}\!\!\left(\frac{\rho_s}{2} \right) \right]^2 } \nonumber \\
&& \qquad \qquad \qquad \quad \times \left| \Psi_{\Phi_{\alpha_\mathrm{min}}} 
		\left[ \frac{(2 k + 1) <\!\!\Phi_{\mathrm{NL}}\!\!> }{ \rho_s + \frac{1}{2}} \right]  \right|^2. \qquad
\label{BerProxDPSKLC}
\end{eqnarray}

\begin{figure}
\centerline{\includegraphics[width = 0.33 \textwidth]{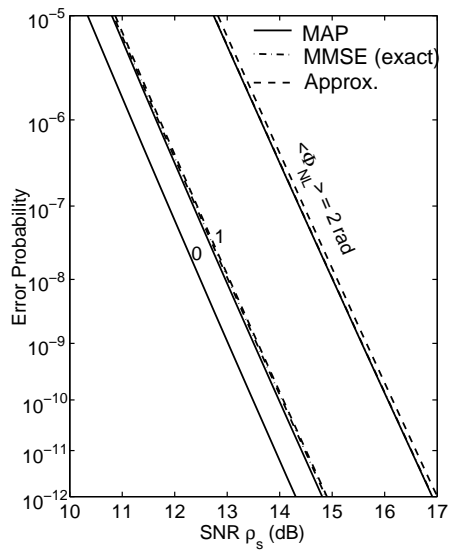}}
\caption{With linear compensation, the error probability of DPSK signal as a function of SNR $\rho_s$.}
\label{figBerDPSKLC}
\end{figure}

\fig{figBerDPSKLC} shows the exact \eqn{BerDPSKLC} and approximated \eqn{BerProxDPSKLC} error probabilities as a function of SNR $\rho_s$ for DPSK signal with linear MMSE compensator.
The scale factor in \eqn{lambdakw} is also numerically optimized to find the linear MAP compensator to minimize the exact error probability \eqn{BerDPSKLC}. 
From \fig{figBerDPSKLC}, the linear MMSE and MAP compensators do not have big difference for linearly compensated DPSK signal.
\fig{figBerDPSKLC} also plots the error probability without nonlinear phase noise.
The approximated error probability in \fig{figBerDPSKLC} is the same as that in \cite{ho0309b} but calculated for large number of fiber spans.
From \fig{figBerDPSKLC}, the approximated error probability \eqn{BerProxDPSKLC} always slightly overestimates the error probability.

\begin{figure}
\centerline{\includegraphics[width = 0.4 \textwidth]{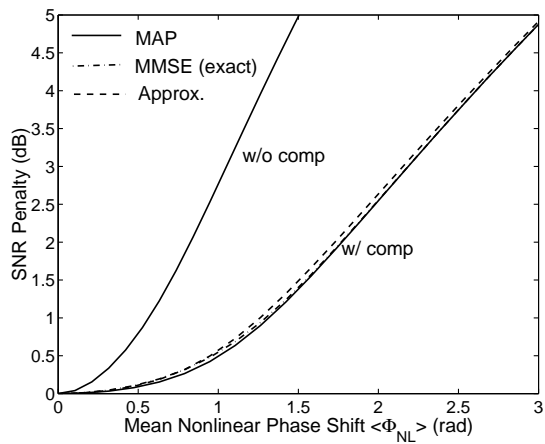}}
\caption{With linear compensation, the SNR penalty of DPSK signal as a function of mean nonlinear phase shift $<\!\!\Phi_{\mathrm{NL}}\!\!>$.}
\label{figPenDPSKLC}
\end{figure}

\fig{figPenDPSKLC} shows the SNR penalty of DPSK signal for an error probability of $10^{-9}$ calculated by the  exact \eqn{BerDPSKLC} and approximated \eqn{BerProxDPSKLC} error probability formulae for DPSK signal with linear MMSE and MAP compensation.
The SNR penalty given by the approximated error probability \eqn{BerProxDPSKLC} is the same as that in \cite{ho0309b} but for large number of fiber spans.
The discrepancy between the exact and approximated error probability with either MMSE or MAP criteria is very small.
The largest discrepancy between the exact and approximated SNR penalty with MMSE criterion is about 0.1 dB at a mean nonlinear phase shift $<\!\!\Phi_{\mathrm{NL}}\!\!>$ of $1.74$ rad. 
Both with the exact error probability \eqn{BerDPSKLC}, the linear MMSE and MAP compensators have the largest discrepancy of 0.06 dB at a mean nonlinear phase shift $<\!\!\Phi_{\mathrm{NL}}\!\!>$ of $0.95$ rad. 

\fig{figPenDPSKLC} also shows the SNR penalty of DPSK signal without compensation from \eqn{figPenDPSK} using the exact error probability formula there.
Similar to \cite{ho0309b}, for the same SNR penalty, DPSK signal with linear compensator can tolerate slightly larger than twice the mean nonlinear phase noise without linear compensation. 

For a power penalty less than 1 dB, the mean nonlinear phase shift  $<\!\!\Phi_{\mathrm{NL}}\!\!>$ must be less than 1.31 rad, slightly larger than twice that of \fig{figPenDPSK} of 0.57 rad without compensation. 
The optimal operating level of the mean nonlinear phase shift  $<\!\!\Phi_{\mathrm{NL}}\!\!>$ is about 1.81 rad such that the increase of power penalty is always less than the increase of mean nonlinear phase shift,  slightly smaller than twice that of \fig{figPenDPSK} of 1 rad without compensation.

\section{Nonlinear Compensation of Nonlinear Phase Noise}
\label{sec:nlcomp}

This section finds the optimal MAP and MMSE compensators for a PSK signal with nonlinear phase noise. 
We will first derive the joint p.d.f. of received amplitude and phase of $p_{R, \Phi_r|\theta_0}(r, \theta)$ given $\theta_0$ is the transmitted phase.
The MAP detector is derived through the Neyman-Pearson criterion \cite[Sec. 5.3]{mcdonough2}.
The error probability is then calculated based on the p.d.f. of $p_{R, \Phi_r|0}(r, \theta)$.
Based on the characteristic function of nonlinear phase nosie condition on the received intensity, the optimal MMSE detector for signals with nonlinear phase noise is also derived analytically.

\subsection{Joint Distribution of Received Amplitude and Phase}

The received phase of \eqn{PhiR} is confined to the range of $[-\pi, +\pi)$. 
The joint PDF of received amplitude and phase $p_{R, \Phi_r|\theta_0}(r, \theta)$ can be modeled as a periodic function of $\theta$ with a period of $2 \pi$ and expanded as a Fourier series as

\begin{eqnarray}
p_{R, \Phi_r|\theta_0}(r, \theta) = \frac{p_R(r)}{2 \pi} + \frac{1}{\pi} 
\sum_{m =1}^{+\infty}\Re \! \left\{ C_{m}(r) e^{ j m (\theta - \theta_0)} \right\}, \nonumber \\
r \geq 0, \qquad \qquad
\label{pdfRPhiR}
\end{eqnarray}

\noindent where $p_R(r)$ is the PDF of received amplitude of \eqn{pdfR}, $C_m(r)$ is the $m$th Fourier coefficient as a function of the received amplitude $r$, and $\Re \!\{\cdot\}$ denotes the real part of a complex number.
The PDF of \eqn{pdfRPhiR} has been simplified using the relationship of $C_{-m}(r) = C^*_m(r)$. 
It is also obvious that $\int_{-\pi}^{+\pi} p_{R, \Phi_r|\theta_0}(r, \theta) \ud \theta = p_R(r)$.

As the Fourier series of the PDF $p_{R, \Phi_r|\theta_0}(r, \theta)$ \eqn{pdfRPhiR}, the Fourier coefficients of $C_m(r)$ are the Fourier transform of the PDF \eqn{pdfRPhiR} with respect to $\theta$ at integer ``angular frequency'' of $m$.
The Fourier transform of a PDF is its characteristic function and $C_m(r)$ can come from the ``partial'' Fourier transform of the joint PDF of nonlinear phase noise, received intensity, and phase of amplifier noise.
The Fourier coefficients are \cite{huang03a}

\begin{equation}
C_m(r) = q_{\Phi, R, \Theta_n}^{*}\left(- m 	\frac{ <\!\! \Phi_{\mathrm{NL}}\!\!>} {\rho_s + 1/2}, r, m  \right),
\label{CmR1}
\end{equation}

\noindent where $q_{\Phi, R, \Theta_n}(\nu, r, m)$ cannot be found directly but

\begin{equation}
q_{\Phi, Y, \Theta_n}(\nu, y, m) = \mathcal{F}^{-1}_{\omega} \left\{ \Psi_{\Phi, Y, \Theta_n} \right\}(\nu, y, m)
\end{equation}

\noindent given by \eqn{cfPhiThetampdfY} with $\Psi_{\Phi, Y, \Theta_n}(\nu, \omega, m)$ given by \eqn{cfPhiYThetam} and $\theta_0 = 0$.
With a change of random variable of $Y = R^2$, the partial PDF and characteristic function of \eqn{cfPhiThetampdfY} becomes

\begin{equation}
  q_{\Phi, R, \Theta_n}(\nu, r, m)
   	=   \frac{r \Psi_\Phi(\nu) }{  \sigma_\nu^2} \exp \!\! \left( -\frac{r^2 + | \vec{\xi}_\nu|^2  }{2 \sigma_\nu^2} \right) \!\!
        I_m \left( \frac{r | \vec{\xi}_\nu|}{\sigma_\nu^2} \right) 
\label{cfPhiThetampdfR}
\end{equation}

\noindent for $m \geq 0$ and $ q_{\Phi, R, \Theta_n}(\nu, r, -m) = q_{\Phi, R, \Theta_n} (\nu, r, m)$.

Based on \eqn{CmR1}, we obtain

\begin{equation}
C_m(r) = \frac{ r \Psi_m}{s_m} \exp\left( -\frac{r^2 + \alpha_m^2  }{2 s _m} \right) 
        I_m \left( \frac{ \alpha_m r }{s_m} \right),\quad m \geq 1,
\label{CmR}
\end{equation}

\noindent with 

\begin{equation}
\Psi_m = \Psi_\Phi \left(m \frac{ <\!\! \Phi_{\mathrm{NL}}\!\!>} {\rho_s + 1/2} \right),
\end{equation}

\begin{equation}
\alpha_m = \sqrt{\rho_s} \sec \left[ \left( jm \frac{ <\!\! \Phi_{\mathrm{NL}}\!\!>} {\rho_s + 1/2} \right)^{\frac{1}{2}} \right], 
\end{equation}

\noindent and

\begin{equation}
s_m =   \frac{1}{2}\left( jm \frac{ <\! \Phi_{\mathrm{NL}}\!>} {\rho_s + 1/2} \right)^{-\frac{1}{2}} \tan \left[ \left( jm \frac{ <\! \Phi_{\mathrm{NL}}\!>} {\rho_s + 1/2} \right)^{\frac{1}{2}} \right].
\end{equation}

\noindent where $\Psi_\Phi(\nu)$ is the characteristic function of nonlinear phase noise of \eqn{cfPhi}.
Using method in quantum field theory, a PDF similar to \eqn{pdfRPhiR} was derived in \cite{turitsyn03}.  

\begin{figure}
\centerline{\includegraphics[width = 0.35 \textwidth]{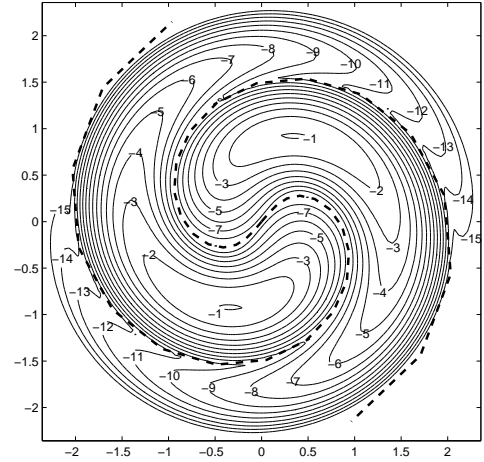}}
\caption{The distribution of received electric field for binary PSK signal.
The contour lines are in logarithmic scale and both $x$ and $y$ axes are normalized to unity mean amplitude.
The dashed line is the decision boundary.
The SNR is  $\rho_s = 18$ and the mean nonlinear phase shift is $<\!\! \Phi_{\mathrm{NL}}\!\!> = 2$ rad.}
\label{figpdfXY}
\end{figure}

For a PSK signal with $\theta_0 \in \{0, \pi\}$, \Fig{figpdfXY} shows the distribution of the received electric field similar to \fig{figyinyang}(b).
The contour lines of \fig{figpdfXY} are the logarithmic of the PDF of $\frac{1}{2}\left[ p_{R, \Phi_r|0}(r, \theta) + p_{R, \Phi_r|\pi}(r, \theta) \right]$ after the conversion to rectangular coordinate. 
The SNR of $\rho_s = 18$ is chosen for an error probability of $10^{-9}$ for binary PSK signal without nonlinear phase noise.
The mean nonlinear phase shift of $<\!\! \Phi_{\mathrm{NL}}\!\!> = 2$ rad is chosen to about double the optimal operating point estimated in \cite{gordon90, ho0309b}.

Similar to those of \fig{figyinyang}, the helix shape distribution of \fig{figpdfXY} clearly shows that the rotation due to nonlinear phase noise is correlated with the amplitude (the distance to the origin). 
The optimal MAP detector is derived here.

\subsection{Optimal MAP Detector}

For a PSK signal with $\theta_0 \in \{0, \pi\}$, using the Neyman-Peason lemma with unity likelihood ratio \cite[Sec. 5.3]{mcdonough2}, the optimal decision regions of the MAP detector are given by

\begin{eqnarray}
\mathcal{R}_0 &=& \Big\{ r, \theta \ \big| \  p_{R, \Phi_r|0}(r, \theta) \geq p_{R, \Phi_r|\pi}(r, \theta) \Big\}, \label{region0} \\
\mathcal{R}_1 &=& \Big\{ r, \theta \ \big| \  p_{R, \Phi_r|\pi}(r, \theta) > p_{R, \Phi_r|0}(r, \theta) \Big\}  
\label{region1}
\end{eqnarray}

\noindent for $\theta_0 = 0$ and $\theta_0 = \pi$, respectively.
The error probability is

\begin{eqnarray}
p_e & = & \frac{1}{2} \int_{\mathcal{R}_0}  p_{R, \Phi_r|\pi}(r, \theta) \ud r \ud \theta 
          + \frac{1}{2} \int_{\mathcal{R}_1}  p_{R, \Phi_r|0}(r, \theta) \ud r \ud \theta \nonumber \\
   & = & \int_{\mathcal{R}_0}  p_{R, \Phi_r|\pi}(r, \theta) \ud r \ud \theta  \nonumber \\
   & = & \int_{\mathcal{R}_1}  p_{R, \Phi_r|0}(r, \theta) \ud r \ud \theta.
\label{BerMAP1}
\end{eqnarray}

The decision regions of $\mathcal{R}_0$ and $\mathcal{R}_1$ have a boundary of $\theta_c(r) \pm \pi/2, r \geq 0$, where $\theta_c(r)$ is the center phase of the decision regions of $\mathcal{R}_0$.
With the joint PDF of \eqn{pdfRPhiR}, the error probability for MAP detector of \eqn{BerMAP1} becomes

\begin{equation}
p_e = \frac{1}{2} - \frac{2}{\pi} 
  \sum_{k = 0}^{+\infty} \frac{(-1)^k}{2 k + 1} \int_{0}^{\infty}
      \Re \!\left\{ C^*_{2 k +1}(r) e^{j(2k+1)\theta_c(r)} \right\} \ud r.
\label{BerMAP}
\end{equation}

Using only $p_{R, \Phi_r|0}(r, \theta)$, because $p_{R, \Phi_r|\pi}(r, \theta) = p_{R, \Phi_r|0}(r, \theta-\pi)$, the center phase of $\theta_c(r)$ can be determined by

\begin{equation}
p_{R, \Phi_r|0}\left(r, \theta_c(r) + \frac{\pi}{2}\right) = p_{R, \Phi_r|0}\left(r, \theta_c(r) - \frac{\pi}{2}\right).
\label{ThetaCr}
\end{equation}

\Fig{figpdfXY} also shows the decision boundary of $\theta_c(r) \pm \pi/2$ based on \eqn{region0} and \eqn{region1} with center phase given by \eqn{ThetaCr}.
The decision boundary is the ``valley'' between two peaks of the PDF of $p_{R, \Phi_r|0}(r, \theta)$ and $p_{R, \Phi_r|\pi}(r, \theta)$, respectively.

As discussed earlier, the optimal MAP detector can be implemented based on the decision boundary of \fig{figpdfXY}, resemble to the Yin-Yang logo of Chinese mysticism, called the Yin-Yang detector in \cite{ho0403a}. 
The same MAP detector can also be implemented as a nonlinear compensator by adding an angle of $\theta_c(r)$ from \eqn{ThetaCr} to the received phase.
The Yin-Yang detector is equivalent to the nonlinear compensator.

In this paper, the optimal MAP detector is derived based on $\theta_c(r)$ as a function of the received amplitude.
Because the relationship between intensity and amplitude is a monotonic function of $Y = R^2, R \geq 0$, the nonlinear phase noise is considered to be compensated by the received intensity instead of the received amplitude.

The more popular DPSK signals can also be compensated using $\theta_c(r_1)-\theta_c(r_2)$, where $r_1$ and $r_2$ are the received amplitudes in two consecutive symbols.
While the compensator for DPSK signals can be constructed using the center phase of \eqn{ThetaCr}, the evaluation of the error probability of nonlinearly compensated DPSK signals is difficult.

\subsection{Optimal MMSE Detector}

The optimal MMSE compensator estimates the normalized nonlinear phase using the received intensity by the conditional mean of $E\{\Phi | R\} = E\{ \Phi | Y\} = E\{ \Phi | R^2\}$ \cite{ho0306}.
The conditional mean is the Bayes estimator that minimizes the variance of the residual nonlinear phase noise without the constraint of linearity \cite[Sec. 10.2]{mcdonough2}.
We call the estimation of $E\{ \Phi | R\}$ an estimation based on received intensity although it is actually based on received amplitude. 
To find the conditional mean $E\{ \Phi | R\}$, either the conditional PDF~of $p_{\Phi | R}(\theta)$ or the conditional characteristic function of  $\Psi_{\Phi | R}(\nu)$ is required.
The conditional characteristic function of $\Psi_{\Phi | R}(\nu)$ can be found using the relationship of 

\begin{equation}
\Psi_{\Phi | Y}(\nu) = \frac{q_{\Phi, Y}(\nu, y)}{ p_Y(y)},
\label{cfPhi|Y}
\end{equation} 

\noindent where $q_{\Phi, Y}(\nu, y)$ \eqn{cfPhipdfY} is the partial characteristic function of normalized nonlinear phase noise and PDF of received intensity and $p_Y(y)$ \eqn{pdfY} is the PDF of received intensity.
Using \eqn{cfPhipdfY} and \eqn{pdfY}, we obtain

\begin{equation}
\Psi_{\Phi | Y}(\nu)  =  \frac{\Psi_\Phi(\nu) \exp\left[ -\frac{y + | \vec{\xi}_\nu|^2  }{2 \sigma_\nu^2} \right] 
I_0\left(\sqrt{y} \frac{|\vec{\xi}_\nu|} { \sigma_\nu^2} \right) }
{ 2 \sigma_\nu^2 \exp\left[ -(y +\rho_s  ) \right] 
I_0\left(2 \sqrt{y\rho_s}  \right) }.
\end{equation}

The optimal MMSE compensator is the conditional mean of the normalized nonlinear phase noise given the received intensity of $Y = R^2$, we obtain

\begin{eqnarray}
E\{ \Phi | R\} & =& -j \left. \frac{d}{d \nu}  \Psi_{\Phi | Y}(\nu|y) \right|_{\nu =0, y = r^2} \nonumber \\
 & = & \frac{1}{6} + \frac{1}{3} \rho_s + \frac{1}{3} r^2 + \frac{\sqrt{\rho_s} r }{3} 	
	\frac{I_1(2r \sqrt{\rho_s} )}{I_0(2r \sqrt{\rho_s} )}.
\label{optcomp}
\end{eqnarray}

\noindent Other than the normalization, the optimal MMSE compensator of \eqn{optcomp} is similar to that of \cite{ho0306} for finite number of fiber spans.
The optimal MMSE compensator \eqn{optcomp} depends solely on the system SNR.

The normalized residual nonlinear phase noise is $\phi_e =  \phi -E\{ \Phi | R\}$,  its conditional variance is

\begin{eqnarray}
\lefteqn{\sigma_{\Phi_e|R}^2(r) =  - \left. \frac{d^2}{d \nu^2}  \Psi_{\Phi | Y}(\nu|y) \right|_{\nu =0, y = r^2} 
	- E\{ \Phi | R\} ^2} \nonumber \\
&&\qquad =  \frac{1 + 4 (\rho_s + r^2)}{90}  + \frac{ r^2 \rho_s} {9} \nonumber \\
&&\qquad \quad	+ \frac{ \sqrt{\rho_s} r}{45}  \frac{I_1(2r \sqrt{\rho_s})}{I_0(2r  \sqrt{\rho_s} )} 	
	- \frac{r^2\rho_s}{9} \frac{I^2_1(2r \sqrt{\rho_s})}{I^2_0(2r  \sqrt{\rho_s} )}. 
\end{eqnarray}

The variance of the normalized residual nonlinear phase noise can be numerically integrated as

\begin{equation}
\sigma_{\Phi_e}^2= \int_0^\infty \sigma_{\Phi_e|R}^2(r) p_R(r) \ud r,
\end{equation}

\noindent where $p_R(r)$ is the PDF~of the received amplitude of \eqn{pdfR}.

Similar to the error probability of \eqn{BerMAP}, the error probability for MMSE detector is

\begin{eqnarray}
p_e & = & \frac{1}{2} - \frac{2}{\pi} 
  \sum_{k = 0}^{+\infty} \frac{(-1)^k}{2 k + 1} \! \int_{0}^{\infty}
      \Re \!\Bigg\{ C^*_{2 k +1}(r) \nonumber \\
  &&  \times  \exp\left[ \frac{  j(2k+1)  <\!\! \Phi_{\mathrm{NL}}\!\!>  }{\rho_s +1/2} E\{\Phi|R\} \right] \Bigg\} \ud r. \nonumber \\
\label{BerMMSE}
\end{eqnarray}

\noindent The factor of $ <\!\! \Phi_{\mathrm{NL}}\!\!>\!\!\!/(\rho_s + 1/2)$ scales the normalized nonlinear phase noise $\Phi$ in \eqn{optcomp} to the actual nonlinear phase noise $\Phi_{\mathrm{NL}}$.

\begin{figure}
\centerline{\includegraphics[width = 0.4 \textwidth]{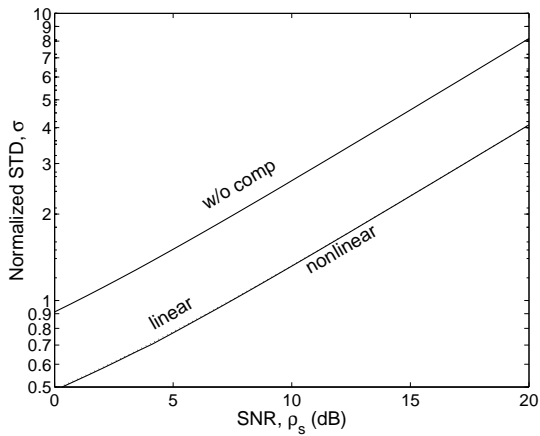}}
\caption{The STD of normalized nonlinear phase noise with and without compensation. 
The STD with linear and nonlinear compensator are showed as dotted- and solid lines, respectively.
}
\label{figsigma}
\end{figure}

\Fig{figsigma} plots the standard deviation (STD) of the normalized nonlinear phase noise with and without compensation as a function of SNR $\rho_s$.
The STD of residual nonlinear phase noise with linear and nonlinear compensator are shown as almost overlapped solid and dotted lines, respectively.
The STD of nonlinear compensator is about $0.2 \%$ less than that of linear compensator.
\Fig{figsigma} confirms the results of \cite{ho0306} that linear and nonlinear MMSE compensator performs the same in term of the variance of residual nonlinear phase noise.

\subsection{Numerical Results}

\begin{figure}
\centerline{\includegraphics[width = 0.4 \textwidth]{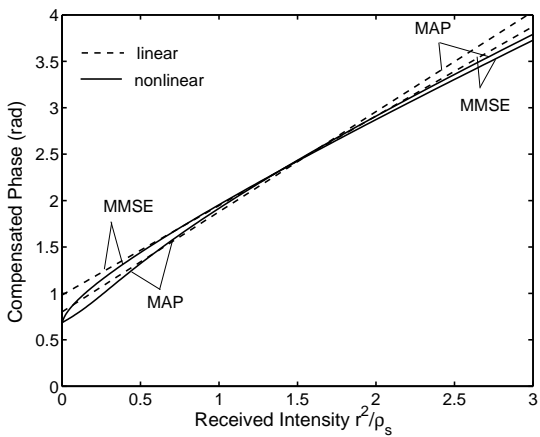}}
\caption{
The compensated phase as a function of received intensity.
System parameters are the same as that in \fig{figpdfXY}.
}
\label{figctrphase}
\end{figure}

\Fig{figctrphase} plots the center phase of $\theta_c(r)$ from \eqn{ThetaCr} as a function of the received amplitude of $r^2$. 
The system parameters of \fig{figctrphase} are the same as that of \fig{figpdfXY}.
As discussed earlier, the center phase of Yin-Yang detector is the same as the compensated phase of a compensator.
The center (or compensated) phase of the nonlinear MMSE compensator of \eqn{optcomp} is also plotted in \fig{figctrphase} for comparison.
The phase of \eqn{optcomp} is scaled by $<\!\!\Phi_\mathrm{NL}\!\!>/(\rho_s + \frac{1}{2})$.
The compensated phases of the linear compensator designed by MMSE or MAP criteria of Sec. \ref{sec:lincomp} are also plotted in \fig{figctrphase} as dashed-lines for comparison.
In \fig{figctrphase}, the received intensity is normalized with respect to the SNR of $\rho_s$. 

From \fig{figctrphase}, nonlinear MMSE or MAP compensated curves are very close to the linear compensated curves, especially when the received intensity is near its mean value of about $r^2/\rho_s = 1$. 
When the STD of \fig{figsigma} is evaluated, the STD is mainly contributed from the region where the random variable is close to its mean value.
Compared the linear and nonlinear compensated phases of \fig{figctrphase}, the STD of \fig{figsigma} for linear and nonlinear MMSE compensator should not have significant difference.
From \fig{figsigma}, the linear and nonlinear MAP compensated phases are also very close to each other. 
In the region closes to $r^2/\rho_s = 1$, because the nonlinear MAP compensated phase is closer to the linear MAP compensated phase than the nonlinear MMSE compensated phase, the linear MAP compensated phase has better performance than the nonlinear MMSE compensated phase. 

\begin{figure}
\centerline{\includegraphics[width = 0.35 \textwidth]{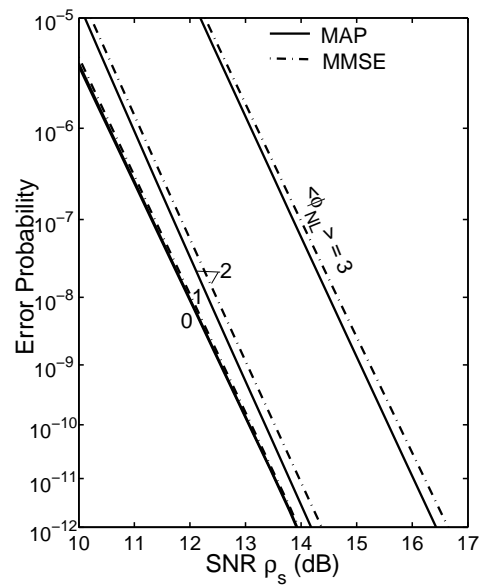}}
\caption{Error probability of a PSK signal with optimal MAP or MMSE detector.
}
\label{figBerPSKNL}
\end{figure}

\Fig{figBerPSKNL} shows the error probability given by \eqn{BerMAP} and \eqn{BerMMSE} for optimal MAP and MMSE detectors, respectively, as a function of SNR $\rho_s$.
\Fig{figBerPSKNL} also plots the error probability of a PSK signal without nonlinear phase noise of $p_{e} = \frac{1}{2}\mathrm{erfc} \sqrt{\rho_s}$.
From \fig{figBerPSKNL}, the optimal MMSE detector does not minimize the error probability.
The optimal compensated phase of \eqn{ThetaCr} always gives a smaller error probability that the compensated phase of \eqn{optcomp}.

\begin{figure}
\centerline{\includegraphics[width = 0.4 \textwidth]{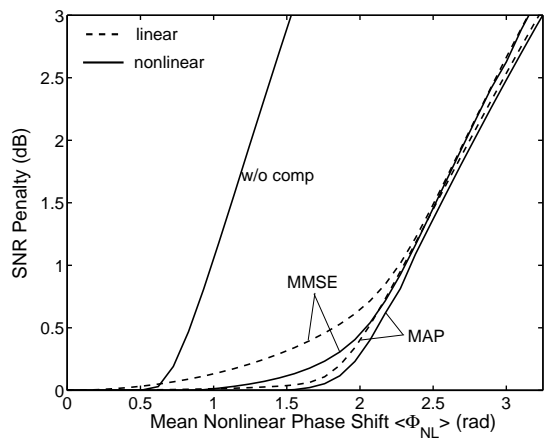}}
\caption{The SNR penalty of PSK signal as a function of mean nonlinear phase shift $<\!\!\Phi_{\mathrm{NL}}\!\!>$.}
\label{figPenPSKNL}
\end{figure}

\fig{figPenPSKNL} shows the SNR penalty of PSK signal for an error probability of $10^{-9}$ calculated by the MAP \eqn{BerMAP} and MMSE \eqn{BerMMSE} error probability formulas.
The SNR penalty with linear MMSE and MAP compensator of \fig{figPenPSKLC} is also plotted as dashed lines for comparison.
The SNR penalty without compensation \fig{figPenPSK} is also shown for comparison.

From \fig{figPenPSKNL}, the nonlinear MMSE compensator performs up to 0.23 dB better than the linear MMSE compensator. 
Although \fig{figsigma} shows that linear and nonlinear MMSE compensators performs almost the same in term of the STD of residual nonlinear phase noise, the error probability has significant difference.
The comparison between \figs{figsigma} and \ref{figPenPSKNL} shows that the variance does not correlate well with the error probability.
From \fig{figPenPSKNL}, the optimal linear MAP detector performs very close to the optimal nonlinear MAP detector. 
The optimal nonlinear MAP detector performs only up to 0.14 dB better than the linear MAP detector. 

\begin{table*}
\caption{PSK Signals with Nonlinear Phase Noise Compensation}
\label{tab:comp}
\begin{center}
\begin{tabular}{|l|l|c|c|c|}
\hline
  &  & \multicolumn{2}{c|}{Mean Nonlinear Phase Shift $<\!\!\Phi_{\mathrm{NL}}\!\!>$} & Max. Diff. \\
\cline{3-4}
  &  & 1-dB Penalty (rad) & Optimal Point (rad) & to Nonlin. MAP (dB) \\
\hline
\multicolumn{2}{|l|}{Without Compensation} & 1.00 & 1.25 & ---- \\
\hline
 MMSE & Linear  & 2.26 & 2.30 & 0.41 \\
    & Nonlinear & 2.31& 2.20& 0.21 \\
\hline
 MAP & Linear &  2.30 & 2.15 & 0.14 \\
    & Nonlinear & 2.35 & 2.12 & 0.00\\
\hline 
\end{tabular}
\end{center}
\end{table*}

Table \ref{tab:comp} shows the mean nonlinear phase shift corresponding to a system with 1-dB SNR penalty and the optimal operating point.
The optimal operating point is found by the condition that the increase of SNR penalty is less than the increase of SNR that is proportional to the mean nonlinear phase shift.
Because of the steepness of the slope, the optimal MAP detector actually gives smaller optimal operating point than other compensation schemes.

\section{Conclusion}
\label{sec:end}

The joint characteristic functions of the nonlinear phase noise with electric field, received intensity, and the phase of amplifier noise are derived analytically the first time.
The nonlinear phase noise is modeled asymptotically as a distributed process for large number of fiber spans.
Replacing the span by span summation of the nonlinear phase noise by an integration, the distributed assumption is valid if the number of fiber spans is larger than 32.
Using the joint characteristic function of the nonlinear phase noise with the phase of amplifier noise, the error probabilities of PSK and DPSK signal are calculated as a series summation.

For PSK signals, the optimal decision region is not centered with respect to the mean nonlinear phase shift.
The dependence between linear and nonlinear phase noise increases the error probability of the signals.
When the received intensity is used to compensate the nonlinear phase noise, based on the principle to minimize the variance of the residual nonlinear phase noise, the optimal linear and nonlinear compensators are derived analytically using the joint characteristic function of the nonlinear phase noise with the received intensity.
Using the exact error probability of systems with linearly compensated nonlinear phase, linear MAP compensator is calculated based on numerical optimization.
Using the distribution of a received signal with nonlinear phase noise, the optimal MAP detector is derived for a phase-modulated signal to minimize the error probability.
The error probability of nonlinear MAP and MMSE detector is also evaluated using the distribution of the received signal.

Having the same variance of residual nonlinear phase noise, nonlinear MMSE compensator performs up to 0.23 dB better than linear MMSE compensator.
The optimal nonlinear MAP compensator performs very close to the optimal linear MAP compensator with a difference less than 0.14 dB.
While the MMSE criterion does not provide a minimum error probability, the linear MAP compensator optimized by numerical methods is able to well approximate the optimal nonlinear detector.  
In practice, the optimal detector can be implemented as a Yin-Yang detector or a nonlinear compensator. 


\end{document}